\documentclass{article}

\usepackage{arxiv}

\usepackage[utf8]{inputenc} 
\usepackage[T1]{fontenc}    
\usepackage{lmodern}        
\usepackage{hyperref}       
\usepackage{url}            
\usepackage{booktabs}       
\usepackage{amsfonts}       
\usepackage{nicefrac}       
\usepackage{microtype}      
\usepackage{graphicx}

\title{Multiple imputation of incomplete multilevel data using Heckman
selection models}

\author{
    Johanna Muñoz
    \thanks{Corresponding author j.munozavila@umcutrecht.nl}
   \\
    Julius Center for Health Sciences and Primary Care \\
    University Medical Center Utrecht, Utrecht University \\
   \\
  \texttt{} \\
   \And
    Matthias Egger
   \\
    Institute of Social and Preventive Medicine \\
    University of Bern \\
   \\
  \texttt{} \\
   \And
    Orestis Efthimiou
   \\
    Institute of Social and Preventive Medicine \\
    University of Bern \\
   \\
  \texttt{} \\
   \And
    Vincent Audigier
   \\
    Laboratoire CEDRIC-MSDMA \\
    Conservatoire National des Arts et Métiers \\
   \\
  \texttt{} \\
   \And
    Valentijn M. T. de Jong \thanks{These authors contributed equally}
   \\
    Julius Center for Health Sciences and Primary Care \\
    University Medical Center Utrecht, Utrecht University \\
   \\
  \texttt{} \\
   \And
    Thomas. P. A. Debray\footnotemark[2]
   \\
    Julius Center for Health Sciences and Primary Care \\
    University Medical Center Utrecht, Utrecht University \\
   \\
  \texttt{} \\
  }

\providecommand{\tightlist}{%
  \setlength{\itemsep}{0pt}\setlength{\parskip}{0pt}}

\newlength{\cslhangindent}
\newlength{\csllabelwidth}
\newlength{\cslentryspacingunit} 
  {}%
  {\par}
\newenvironment{CSLReferences}[2] 
 {
  \setlength{\parindent}{0pt}
  \ifodd #1
  \let\oldpar\par
  \def\par{\hangindent=\cslhangindent\oldpar}
  \fi
  \setlength{\parskip}{#2\cslentryspacingunit}
 }%
 {}
\usepackage{calc}

\usepackage{mathtools}
\usepackage{float}
\usepackage{multicol}
\usepackage{graphicx} 
\usepackage{amsmath} 
\usepackage{booktabs}
\usepackage{longtable}
\usepackage{array}
\usepackage{multirow}
\usepackage{wrapfig}
\usepackage{float}
\usepackage{enumitem}
\usepackage{colortbl}
\usepackage{pdflscape}
\usepackage{tabu}
\usepackage{threeparttable}
\usepackage{makecell}
\usepackage{xcolor}
\usepackage{array} \newcolumntype{R}[1]{>{\raggedleft\let\newline\\\arraybackslash\hspace{0pt}}p{#1}}
\begin{document}
\maketitle

\begin{abstract}
Missing data is a common problem in medical research, and is commonly
addressed using multiple imputation. Although traditional imputation
methods allow for valid statistical inference when data are missing at
random (MAR), their implementation is problematic when the presence of
missingness depends on unobserved variables, i.e.~the data are missing
not at random (MNAR). Unfortunately, this MNAR situation is rather
common, in observational studies, registries and other sources of
real-world data. While several imputation methods have been proposed for
addressing individual studies when data are MNAR, their application and
validity in large datasets with multilevel structure remains unclear. We
therefore explored the consequence of MNAR data in hierarchical data
in-depth, and proposed a novel multilevel imputation method for common
missing patterns in clustered datasets. This method is based on the
principles of Heckman selection models and adopts a two-stage
meta-analysis approach to impute binary and continuous variables that
may be outcomes or predictors and that are systematically or
sporadically missing. After evaluating the proposed imputation model in
simulated scenarios, we illustrate it use in a cross-sectional community
survey to estimate the prevalence of malaria parasitemia in children
aged 2-10 years in five subregions in Uganda.
\end{abstract}


\newpage

\hypertarget{introduction}{%
\section{Introduction}\label{introduction}}

Over the past few years, data sharing efforts have substantially
increased, and researchers increasingly often have access to IPD from
large combined datasets derived from electronic health records (EHR) or
from multiple randomized or observable trials (i.e.~in IPD
meta-analysis, IPD-MA). For example, the clinical practice research
datalink (CRPD) (Herrett et al. 2015) is an electronic health record
dataset in the UK, which has been used in a variety of medical research,
such as the evaluation of health policy and drug efficacy. A recent
example of an IPD-MA is the emerging risk factor collaboration(The
Emerging Risk Factors Collaboration 2007), where data were combined from
approximately 1.1 million individuals across 104 observational studies
to investigate associations of cardiovascular diseases with several
predictors. Individuals in these large datasets tend to be clustered in
centres, countries or studies, where they have been subject to similar
healthcare processes. Moreover, clusters may also differ in participant
eligibility criteria, follow-up length, predictor and outcome
definitions, or in the quality of applied measurement methods. Hence,
heterogeneity between clusters with respect to baseline covariates and
outcomes, while the structure of the correlations between these
variables is likely to be different across different clusters.

A usual problem is that such clustered datasets may contain many
incomplete variables. For example, in registry data it is common that
test results are not available for all patients, as the decision to test
may be at the discretion of the primary care physician or because the
patient refuses to undergo testing. It is also possible that variables
are systematically missing across clusters.(Resche-Rigon et al. 2013)
For instance, in an IPD-MA, studies may have collected information on
different variables. Missing values may thus appear for all participants
of a study in the combined dataset. The presence of missing data can
lead to loss of statistical power, imbalance in cluster size, bias in
parameter estimates and therefore to erroneous conclusions as the
analysis could be based on an unrepresentative sample.

To address the presence of missing data, it is important to consider the
missing mechanism for each incomplete variable. Rubin (1976) (Rubin
1976) identified three missing mechanisms where the probability of
missingness: 1) is independent from observed or missing values (missing
completely at random; MCAR), 2) depends on observed data only (missing
at random; MAR), or 3) depends on unobserved information even after
conditioning on all observable variables (missing not at random; MNAR).
Traditional imputation methods are designed to address incomplete data
sets where variables are MCAR or MAR. Their implementation is justified
when there is not systematic difference between units with missing and
with complete data or when the missingness of a variable is strongly
related to variables measured in the study.

Registries are notoriously prone to incomplete variables that are MNAR,
due complex recording processes. (Liu et al. 2021) For example,
laboratory tests are taken only in certain patients based on symptoms
that are often incompletely recorded. Data from randomized trials may
also suffer from MNAR, for example when study participants that
experience unfavorable results drop out of the study. Also,
heterogeneity of the primary objective or resources of the studies may
result in variables relevant to explain the missing process not being
recorded at all in some of the studies.

Modelling data under MNAR mechanism implies to specify information about
the missingness process in addition to assumptions about the observed
data. Two major approach have been used to address MNAR mechanism:
pattern mixture models(Little and Wang 1996) and selection models (Vella
1998). One of the most popular techniques within selection models is the
one proposed by Heckman (1976). (Heckman 1976) Briefly, the Heckman
selection model corrects for selection bias by estimating two linked
equations: a outcome equation, where the missing variable is associated
with predictors, and a selection equation, that accounts for the
inclusion of observations in the sample. An important feature of the
Heckman selection model is that it does not assume data to be MNAR, so
that it can also be used when data are MCAR or MAR. It therefore offers
an appealing solution to incomplete data sets when the missingness
mechanism is not precisely known.

Over the past few years, several extensions and adaptations to the
Heckman selection model have been proposed for multiple imputation.
Among them, Galimard et al.(2016) (Galimard et al. 2016) implemented a
chained equations imputation method for continuous variables, which was
extended to binary and categorical variables by employing copula
estimates. (Galimard et al. 2018) Also, Ogundimu and Collins (2019)
(Ogundimu and Collins 2019) proposed a chained equations imputation
method that is less dependent on normality assumptions.

In clustered data sets, multilevel imputation methods are required to
properly propagate uncertainty within and across clusters. (Audigier et
al. 2018) However, to our knowledge, existing multilevel imputation
methods mainly focus on situations where data are MAR, and do not adopt
Heckman selection models. Although Hammon and Zinn (2020) (Hammon and
Zinn 2020) recently proposed an extension that allows for the inclusion
of random intercept effects, it can only be used for binary missing
variables and assumes that the effect of explanatory variables on the
missingness mechanisms is common across clusters.

Therefore, the aim of this work is to develop a multilevel imputation
method for continuous and binary variables that are both sporadically
and systematically MNAR, and that this can be applied for an incomplete
outcome or multiple incomplete predictors in the data sets.

In section \ref{sec:heckman} we provide an introduction to the Heckman
model and its estimation, and we extend it to a hierarchical setting. In
section \ref{sec:methods} we define the main steps of the proposed
imputation method. In section \ref{sec:simulation} we provide the
settings and results of a simulation study to evaluate the performance
of our imputation method. In section \ref{sec:illustration} we
illustrate the method using the survey information collected in
different sub-districts in Uganda to estimate the prevalence of malaria
in children. Finally, in section \ref{sec:discussion} we summarize the
results, outline limitations and propose future extension of our method.

\hypertarget{the-heckman-model}{%
\section{The Heckman model}\label{the-heckman-model}}

\label{sec:heckman} The Heckman selection model was initially proposed
as a method to correct for selection bias, in which individuals are not
randomly selected from the population, leading to inconsistent estimates
and erroneous conclusions. (Heckman 1976)

Selection bias occurs when the inclusion of an observation into the
sample is influenced by unobserved variables (e.g., the respondent's
level of trust toward healthcare entities may cause them to self-select
out of the study or refuse to sign consent for a test), which in turn
either influence the outcome of interest (e.g., the result of a blood
test), or are related to other unobserved variables that influence the
outcome of interest. (Vella 1998)

\begin{figure}[!htb]
\vspace{-0.7cm}
   \centering
   \includegraphics[scale=0.35]{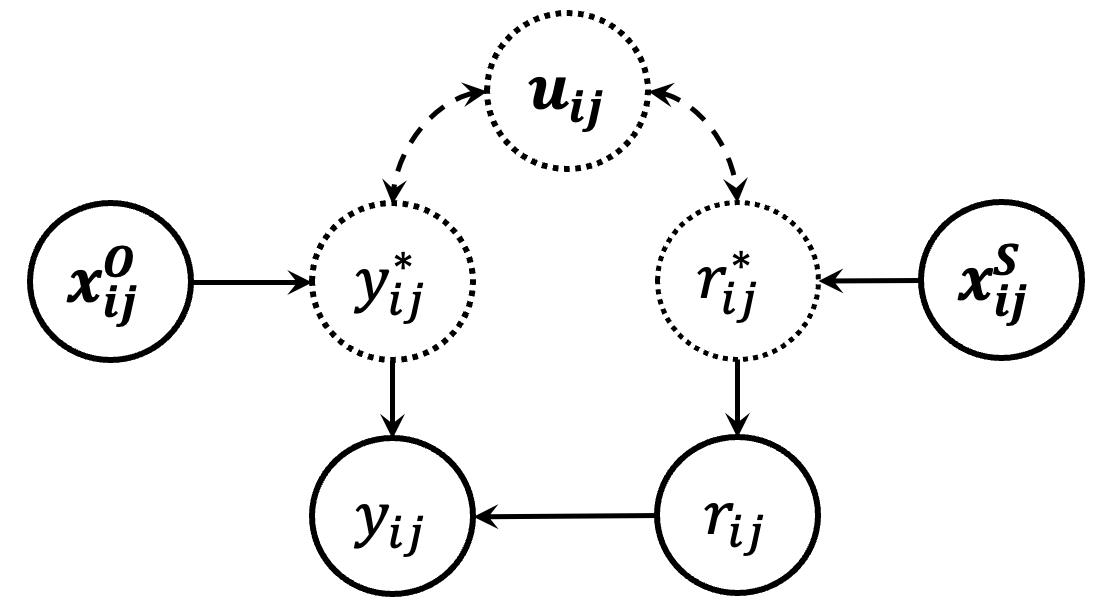}
   \caption{DAG Heckman selection model: here the nodes (dotted lines = latent variables, continuous lines = observed variables) describe the relationship between $y^*_{ij}$ the latent response and $r^*_{ij}$ the latent selection variables of a $j$th unit in a $i$th cluster, that are dependable on $ \boldsymbol{x_{ij}^O}$ and $ \boldsymbol{x_{ij}^S}$ sets of predictors and are correlated through $ \boldsymbol{u_{ij}}$ unobservable variables.}
   \label{fig:DAG}
\end{figure}

This is visualized in Figure \ref{fig:DAG}, where for the \(j\)th
individual or unit within the \(i\)th cluster, there is \(y_{ij}^*\), a
latent outcome variable, and \(r_{ij}^*\), a latent selection variable,
which are correlated through \(\boldsymbol{u_{ij}}\) an unobserved or
unrecorded variable, with \(i\in[1,2,...,N]\) and \(j\in[1,2,...,n_i]\).
Here, both latent variables are related to the sets of predictor
covariates \(\boldsymbol{x_{ij}^O}\) and \(\boldsymbol{x_{ij}^S}\). From
\(r_{ij}^*\) one can derive \(r_{ij}=I(r_{ij}^*>=0)\) a selection
indicator of \(y_{ij}^*\) into the sample, and with this in turn, one
can define \(y_{ij}=y_{ij}^*,\forall r_{ij}=1\) the observable outcome
variable.

Denoting \(\boldsymbol{y^*_i}=(y^*_{i1},y^*_{i2},...,y^*_{in_i})^T\) and
\(\boldsymbol{r^*_i}=(r^*_{i1},r^*_{i2},...,r^*_{in_i})^T\) the vectors
of latent outcomes and latent selections in the cluster \(i\), then
Heckman's model is defined by two main equations: the outcome equation
(\ref{eq:1}), which describes the relation between the latent outcome
(\(y_{ij}^*\)) and a set of covariates
(\(\boldsymbol{X_{i}^O}=(\boldsymbol{x^{O}_{i1}},\boldsymbol{x^{O}_{i2}},...,\boldsymbol{x^{O}_{in_i}})^T\)),
and the selection equation (\ref{eq:2}) which models the likelihood that
the outcome is observed in the sample as a function of another set of
covariates
(\(\boldsymbol{X_{i}^S}=(\boldsymbol{x^{S}_{i1}},\boldsymbol{x^{S}_{i2}},...,\boldsymbol{x^{S}_{in_i}})^T\)).
\begin{align}
\boldsymbol{y^*_{i}}&=\boldsymbol{X^O_{i}}\boldsymbol{\beta^O_{i}}+\boldsymbol{\epsilon^O_{i}}\label{eq:1}\\
\boldsymbol{r^*_{i}}&=\boldsymbol{X^S_{i}}\boldsymbol{\beta^S_{i}}+\boldsymbol{\epsilon^S_{i}}\label{eq:2}
\end{align}

Here \(\boldsymbol{\beta^O_{i}}\) and \(\boldsymbol{\beta^S_{i}}\) are
\(p\times1\) and \(q\times1\) coefficient parameter vectors and
\(\boldsymbol{\epsilon^O_{i}}=(\epsilon^O_{i1},\epsilon^O_{i2},...,\epsilon^O_{in_i})^T\)
and
\(\boldsymbol{\epsilon^S_{i}}=(\epsilon^S_{i1},\epsilon^S_{i2},...,\epsilon^S_{in_i})^T\)
are the residual terms vectors for the outcome and selection equations,
respectively.

Generally the same variables can be used on the matrix of predictor
variables \(\boldsymbol{X^O_{i}}\) and \(\boldsymbol{X^S_{i}}\).
However, to avoid multicollinearity problems(Puhani 1997), it is
recommended to include in \(\boldsymbol{X^S_{i}}\) at least one variable
that is not included in the outcome model. (Angrist and Krueger 2001)
This variable is commonly known as an exclusion restriction variable
(ERV), and should only be associated with the selection in the sample
\(\boldsymbol{r^*_{i}}\) (relevance condition) but not with the actual
observation \(\boldsymbol{y^*_{i}}\) (exclusion condition).(Gomes et al.
2020) The ERV meets by definition the relevance and exclusion conditions
in order to provide independent information about the selection process
and to facilitate the estimation of the Heckman model.

In the presence of selection bias, the aforementioned outcome equation
will yield biased estimates of \(\boldsymbol{\beta^O_{i}}\) if no
efforts are made to adjust for the non-representativeness of the
observed \(\boldsymbol{X^O_{i}}\) and \(\boldsymbol{y_{i}}\) values. For
this reason, the Heckman model aims to jointly estimate the outcome and
selection equation by defining a relation between their respective error
distributions. For instance, Heckman's original model (Heckman 1976)
assumes that residual terms have a bivariate normal distribution (BVN),

\[\begin{pmatrix}
\boldsymbol{\epsilon^O_{i}} \\
\boldsymbol{\epsilon^S_{i}}
\end{pmatrix}\sim N\left(\begin{pmatrix}
0 \\
0
\end{pmatrix},\begin{pmatrix}
\sigma^2_i & \rho_i\sigma_i \\
\rho_i\sigma_i & 1
\end{pmatrix}\right)
\]

where \(\sigma_i\) corresponds to the variance of the error in the
outcome equation and \(\rho_i\) to the correlation between the error
terms of the outcome and selection equations in the \(i\)-th cluster. As
in a probit model, this model assumes a unit variance for the error term
of the selection equation. The unit variance has no consequence on the
observable values of \(r_{ij}=\{0,1\}\), since they only depend on the
sign of \(r^*_{ij}\) and not on its scale.

The interpretation of \(\rho_i\) is fairly straightforward. When
\(\rho_i=0\), the participation does not affect the outcome model and
missing data can be considered MCAR (if data are missing completely at
random) or MAR (if missingness is already explained by
\(\boldsymbol{x_{ij}^O}\)). Conversely, when \(\rho_i\neq 0\), this
suggests that data are MNAR.

\hypertarget{heckman-model-estimation}{%
\subsection{Heckman model estimation}\label{heckman-model-estimation}}

Under the BVN assumption, the parameters of the Heckman model
coefficients can be estimated using the two-step Heckman method (H2S)
(Heckman 1976) or the full information maximum likelihood method
(FIML).(Amemiya 1984) However, both methods may lead to inconsistent
estimators when the true underlying distribution is not the BVN. (Gomes
et al. 2020) To overcome this problem, other approaches have been
proposed that relax the distribution assumptions, among them copula
models. (Smith 2003) The copula approach uses a function, known as a
copula, that joins the marginal distribution of the error terms of the
selection and outcome equation which are specified separately. The
Heckman model can be expressed with the following bivariate normal
copula:
\[ F(\boldsymbol{r^*_{i}},\boldsymbol{y^*_{i}}) =\Phi\left(\boldsymbol{r^*_{i}}-\boldsymbol{X^S_{i}}\boldsymbol{\beta^S_{i}},\frac{\boldsymbol{y^*_{i}}-\boldsymbol{X^O_{i}}\boldsymbol{\beta^O_{i}}}{\sigma_i}, \rho_i\right),\]
where \(\Phi\) is the bivariate standard normal cumulative distribution
function. Thus, to estimate the parameters of the Heckman method, it is
sufficient to specify the marginal distributions of the error terms, and
link them with a suitable copula function. In our imputation method we
estimate the Heckman model using the copula method for non-random
selection (Wojtyś, Marra, and Radice 2018) which is a flexible approach
that allows the specification of different parametric distributions of
the selection and outcome variables and also different types of
dependence structure between the two equations.

\hypertarget{hierarchical-model}{%
\subsection{Hierarchical model}\label{hierarchical-model}}

The Heckman model can be extended to hierarchical settings, i.e., in
cases when individuals or sampling units are nested within clusters, as
is the case in EHR or IPD. This hierarchical complexity must not only be
taken into account in the analysis model, but also when dealing with
missing data, thus requiring imputation models that are congenial to the
analysis model, i.e.~that make the same assumptions about the data.

Different procedures can be adopted to combine information between
clusters; however, in our imputation method we opted for the two-stage
approach that is often used in meta-analyses.(Simmonds et al. 2005) This
is because such an approach is computationally less intensive and could
potentially generate fewer convergence problems in the estimation of the
Heckman hierarchical model compared to other approaches. Our method is
based on the following conditional imputation model, whose parameters
are
\(\boldsymbol{\theta}=\{\boldsymbol{\beta^O},\boldsymbol{\beta^S},\boldsymbol{\Psi^S},\boldsymbol{\Psi^O},(\sigma_i,\rho_i)_{1<=i<=N}\}\)
\begin{align*}
\boldsymbol{y^*_{i}}&=\boldsymbol{X^O_{i}}(\boldsymbol{\beta^O}+b^O_{i})+\boldsymbol{\epsilon^O_{i}}\\
\boldsymbol{r^*_{i}}&=\boldsymbol{X^S_{i}}(\boldsymbol{\beta^S}+b^S_{i})+\boldsymbol{\epsilon^S_{i}}\\
b^O_{i}&\sim N(0,\boldsymbol{\Psi^O})\\
b^S_{i}&\sim N(0,\boldsymbol{\Psi^S})
\end{align*} \[\begin{pmatrix}
\boldsymbol{\epsilon^O_{i}} \\
\boldsymbol{\epsilon^S_{i}}
\end{pmatrix}\sim N\left(\begin{pmatrix}
0 \\
0
\end{pmatrix},\begin{pmatrix}
\sigma^2_i & \rho_i\sigma_i \\
\rho_i\sigma_i & 1
\end{pmatrix}\right)
\]

Briefly, in a first stage we estimate the cluster specific parameters of
the Heckman model
\(\widehat{\boldsymbol{\theta_i}}=\boldsymbol{\{\widehat{\beta^O_i}}, \boldsymbol{\widehat{\beta^S_i}}, \widehat{\sigma_i}, \widehat{\rho_i}\}\),
for all \(i\) between 1 and \(N\) clusters. That is, we estimate
separately for each of the clusters, the parameters of the two equations
in each study, the outcome model (\ref{eq:1}) and the selection model
(\ref{eq:2}) via a copula model.

In a second stage, we estimate random effects multivariate meta-analysis
model for the coefficient parameters. Here \(\boldsymbol{\beta_i^O}\)
and \(\boldsymbol{\beta_i^S}\)parameters are assumed to be drawn
independently and identically from a latent multivariate normal
distribution of parameters with a population means
\(\boldsymbol{\beta^O}\) and \(\boldsymbol{\beta^S}\) and population
variance covariance \(\boldsymbol{\psi^O}\) and \(\boldsymbol{\psi^S}\)
respectively. (Higgins, Thompson, and Spiegelhalter 2009) Here we assume
that \(\sigma_i\) and \(\rho_i\) are random variables that come from a
populational distribution and we estimate random effects univariate
meta-analysis models for these parameters.

\hypertarget{using-the-heckman-model-to-impute-missing-data}{%
\section{Using the Heckman model to impute missing
data}\label{using-the-heckman-model-to-impute-missing-data}}

\label{sec:methods} We follow a similar approach proposed by
Resche-Rigon and White (2018) (Resche-Rigon and White 2018) for
multilevel data imputation, which allows us to impute values in very
common scenarios in IPD, e.g., sporadic and systematic missingness
patterns. Our imputation method was created to impute datasets with a
single missing outcome variable or covariate, but it can also be used to
impute multiple incomplete variables in a dataset, as it can be
implemented under a multivariate imputation by chain equations (MICE)
approach.

In the following subsections, we explain the method when the incomplete
variable is the outcome variable but this method can also be used to
impute any incomplete MNAR predictor variable in the dataset. In the
latter case, in the outcome equation (\ref{eq:1}), the incomplete
predictor variable must be specified as the dependent variable and
\(\boldsymbol{X^O_{i}}\) as the set of covariates associated with it. On
the other hand, in the selection model(\ref{eq:2}),
\(\boldsymbol{r^*_{i}}\) corresponds to the latent variable of selection
of the incomplete predictor variable together with its
\(\boldsymbol{X^S_{i}}\) associated set of predictors.

\hypertarget{univariate-imputation}{%
\subsection{Univariate imputation}\label{univariate-imputation}}

Given an outcome variable \(\boldsymbol{y}=(y_1,y_2,...,y_N)^T\), that
consists of \(y^{miss}_{ij}\) missing and \(y^{obs}_{ij}\) observable
values, we generate independent draws from the posterior predictive
distribution for the missing data, \(y_{ij}^{miss}\), given the
observable data information \(y_{ij}^{obs}\).
\[p(y_{ij}^{miss}|y_{ij}^{obs})=\int_{\boldsymbol{\theta}} p(y_{ij}^{miss}|\boldsymbol{\theta},y_{ij}^{obs})p(\boldsymbol{\theta}|y_{ij}^{obs})\mathrm{d}\boldsymbol{\theta}\]
Here we implicitly assume vague prior distributions for each of the
parameters included in the parameter vector \(\boldsymbol{\theta}\).
Because the integration can be performed computationally by sampling
from the posterior predictive distribution
\(p(\boldsymbol{\theta}|y_{ij}^{obs})\), our imputation method can be
carried out in the following two steps:

\begin{enumerate}
\def\labelenumi{\arabic{enumi}.}
\item
  Draw a \(\boldsymbol{\theta}\) parameter vector,
  \(\boldsymbol{\theta^*}\), from
  \(p(\boldsymbol{\theta}|y_{ij}^{obs})\), their posterior distribution.
\item
  Draw \(y_{ij}^{miss}\) from
  \(p(y_{ij}^{miss}|\boldsymbol{\theta^*})\), their predictive
  distribution for a given \(\boldsymbol{\theta^*}\) vector.
\end{enumerate}

Below we describe each step in depth:

\hypertarget{draw-the-boldsymboltheta-parameter-vector}{%
\subsubsection{\texorpdfstring{Draw the \(\boldsymbol{\theta^*}\)
parameter
vector}{Draw the \textbackslash boldsymbol\{\textbackslash theta\^{}*\} parameter vector}}\label{draw-the-boldsymboltheta-parameter-vector}}

\hypertarget{fit-py_ijobsboldsymboltheta_i-the-heckman-selection-model-at-cluster-level}{%
\paragraph{\texorpdfstring{Fit
\(p(y_{ij}^{obs}|\boldsymbol{\theta_i})\), the heckman selection model
at cluster
level}{Fit p(y\_\{ij\}\^{}\{obs\}\textbar\textbackslash boldsymbol\{\textbackslash theta\_i\}), the heckman selection model at cluster level}}\label{fit-py_ijobsboldsymboltheta_i-the-heckman-selection-model-at-cluster-level}}

Initially, we use the copula method to estimate the set of
cluster-specific parameters,
\(\widehat{\boldsymbol{\theta_i}}=\boldsymbol{\{\widehat{\beta^O_i}}, \boldsymbol{\widehat{\beta^S_i}}, \widehat{\sigma_i}, \widehat{\rho_i}\}\),
using all \(j\) units with observable measurements \(y_{ij}^{obs}\)
within each cluster \(i\). Here we obtain the estimates not only the
parameters' point estimates \(\boldsymbol{\widehat{\theta_i}}\), but
also their corresponding \(\boldsymbol{\widehat{S(\theta_i)}}\)
within-cluster variance-covariance matrix.

\hypertarget{fit-a-meta-analysis-model}{%
\paragraph{Fit a meta-analysis model}\label{fit-a-meta-analysis-model}}

In this step, we pool the parameters \(\boldsymbol{\widehat{\theta_i}}\)
with a random effects meta-analysis model using only the clusters with
observable information, i.e., those with no systematically missing
outcome. In particular, we pool the \(p\) coefficients of the
\(\boldsymbol{\beta^O}\) in the outcome equation and estimate a
multivariate random effects meta-analysis model with them. Similarly we
combine all \(q\) coefficient parameters of the \(\boldsymbol{\beta^S}\)
in the selection equation. Thus, we can denote the study specific
coefficients
\[\boldsymbol{\widehat{\beta^O_i}} = \boldsymbol{\beta^O} +\boldsymbol{ b^O_i} + \boldsymbol{\epsilon'^O_{i}}\]
\[\boldsymbol{\widehat{\beta^S_i}} = \boldsymbol{\beta^S} + b^S_i + \boldsymbol{\epsilon'^S_{i}}\]
using the random effects \(b^O_i \sim N(0,\boldsymbol{\Psi^O})\) and
\(b^S_i\sim N(0,\boldsymbol{\Psi^S})\) with sampling errors
\(\epsilon'^O_{i}\) and \(\epsilon'^S_{i}\).

We assume that \(\sigma_i\) and \(\rho_i\) are random variables coming
from a population distribution, so we can express them as:
\[ log(\sigma_i) \sim N( log(\sigma),\psi^{\sigma})\]
\[ tanh^{-1}(\rho_i)\sim N( tanh^{-1}(\rho),\psi^{\rho})\] For each of
them, we perform a univariate random effects meta-analysis. The model
for \(\sigma_i\) is given by the model:
\[log(\hat\sigma_i) = \sigma + b^{\sigma}_i + \epsilon^{\sigma}_{i}\]
with \(b_i^{\sigma} \sim N(0,\psi^{\sigma})\), and
\(\epsilon_i^{\sigma} \sim N(0,var(log(\hat{\sigma}_i)))\). In the case
of an incomplete binary variable, the \(\sigma_i\) parameter is not
specified, so it is not necessary to perform a random effects
meta-analysis model of \(\sigma_i\) or to include it in any of the
following steps in the imputation model.

The model for \(\rho_i\) is given by:
\[tanh^{-1}(\hat\rho_i) = \rho + b^{\rho}_i + \epsilon^{\rho}_{i}\] with
\(b_i^{\rho} \sim N(0,\psi^{\rho})\), and
\(\epsilon_i^{\rho} \sim N(0,var(tanh^{-1}(\hat{\rho}_i)))\).

\hypertarget{draw-the-marginal-parameters-boldsymboltheta}{%
\paragraph{\texorpdfstring{Draw the marginal parameters
\(\boldsymbol{\Theta}\)}{Draw the marginal parameters \textbackslash boldsymbol\{\textbackslash Theta\}}}\label{draw-the-marginal-parameters-boldsymboltheta}}

From the meta-analysis model, we obtain the marginal estimates
\(\widehat{\boldsymbol{\Theta}}=\boldsymbol{\{\widehat{\beta^O}}, \boldsymbol{\widehat{\beta^S}}, \widehat{\sigma}, \widehat{\rho}\}\)
and the between-cluster variance matrix
\(\widehat{\boldsymbol{\psi}}=\boldsymbol{\{\widehat{\Psi^O}}, \boldsymbol{\widehat{\Psi^S}}, \widehat{\psi^{\sigma}}, \widehat{\psi^{\rho}}\}\),
i.e.~variance-covariance matrix of the random effects, with their
corresponding variance-covariance matrices
\(\boldsymbol{\widehat{S_{\Theta}}}\) and
\(\boldsymbol{\widehat{S_{\psi}}}\), which are used to draw the
\(\boldsymbol{\Theta^*}\) and \(\boldsymbol{\psi^*}\) parameters, from
their posterior distribution as follows: (Resche-Rigon et al. 2013)

\begin{align*}
\boldsymbol{\Theta^*}& \sim N(\widehat{\boldsymbol{\Theta}},\widehat{\boldsymbol{S_{\Theta}}})\\
\boldsymbol{\psi^*}& \sim N(\widehat{\boldsymbol{\psi}},\widehat{\boldsymbol{S_{\psi}}})
\end{align*}

\hypertarget{draw-the-cluster-parameters-boldsymboltheta_i}{%
\paragraph{\texorpdfstring{Draw the cluster parameters
\(\boldsymbol{\theta^*_i}\)}{Draw the cluster parameters \textbackslash boldsymbol\{\textbackslash theta\^{}*\_i\}}}\label{draw-the-cluster-parameters-boldsymboltheta_i}}

We draw the shrunked-cluster-parameters \(\boldsymbol{\theta^*_i}\) for
each \(i\) cluster from the following posterior distribution conditional
on \(\boldsymbol{\Theta^*}\) and \(\boldsymbol{\psi^*}\).

\begin{align*}
\theta^*_i\sim N\left(\left({\boldsymbol{\psi^*}}^{-1}+{\boldsymbol{\widehat{S_{\theta_i}}}}^{-1}\right)^{-1}\left(\boldsymbol{\psi^*}^{-1}\boldsymbol{\Theta^*}+\boldsymbol{\widehat{S_{\theta_i}}}^{-1}\boldsymbol{\widehat{\theta_i}} \right),\left({\boldsymbol{\psi^*}}^{-1}+{\boldsymbol{\widehat{S_{\theta_i}}}}^{-1}\right)^{-1} \right)
\end{align*}

As can be seen, the mean and variance of the posterior distribution is a
combination between the estimated marginal and cluster-specific
parameters. Here the weights on the cluster-specific parameters
\(\boldsymbol{\widehat{\theta_i}}\) and the marginal parameters
\(\boldsymbol{\Theta^*}\) are inversely proportional to the within
cluster variance \(\boldsymbol{\widehat{S_{\theta_i}}}\) and between
clusters variance \(\boldsymbol{\psi^*}\). For example, when
\(\boldsymbol{\widehat{S_{\theta_i}}} < \boldsymbol{\psi^*}\) the mean
of the conditional distribution gives more weight to the estimated
cluster-specific parameter. Conversely, when
\(\boldsymbol{\widehat{S_{\theta_i}}}>\boldsymbol{\psi^*}\), more weight
is given to the estimated marginal parameters. Therefore, in case of a
cluster with systematic missingness, it is as if the within-cluster
variance is infinite (\(\boldsymbol{\widehat{S_{\theta_i}}}\to\infty\)),
so all the weight is assigned to the parameters estimated at the
marginal level. That is, when there is no information in a cluster, we
rely entirely on the marginal information.

\hypertarget{draw-y_ijmiss-observation}{%
\subsubsection{\texorpdfstring{Draw \(y_{ij}^{miss}\)
observation}{Draw y\_\{ij\}\^{}\{miss\} observation}}\label{draw-y_ijmiss-observation}}

Having \(\boldsymbol{\theta^*_i}\), the shrunk-cluster parameters vector
for each cluster, we back-transform \(\sigma^*\) and \(\rho^*\) to the
original scale. Then \(y^{miss}_{ij}\), the missing values, can be drawn
from \(p(y^{miss}_{ij}|\boldsymbol{\theta^*_i})\), their predictive
distribution given \(\boldsymbol{\theta^*_i}\), as follows:

\hypertarget{continuous-missing-variable}{%
\paragraph{Continuous missing
variable}\label{continuous-missing-variable}}

The imputed value of \(y_{ij}^{miss}\) can be drawn from the conditional
expectation of \(y_{ij}\) on unobserved measurements: (Greene 2018)
\begin{align*}
\mu&= E[y_{ij}|r_{ij}=0,{\boldsymbol{\beta^O_{i}}^*},\boldsymbol{{\beta^{S}_{i}}^*},{\rho_{i}}^*,{\sigma_i}^*]\\
\mu&= \boldsymbol{x^O_{ij}}\boldsymbol{{\beta^O_{i}}^*}+\rho_{i}^*\sigma_i^*\frac{-\phi(\boldsymbol{x^{S}_{ij}}\boldsymbol{{\beta^{S}_{i}}^*})}{\Phi(-\boldsymbol{x^{S}_{ij}}{\boldsymbol{\beta^{S}_i}^*})}\\
y_{ij}^{miss}&\sim N(\mu,{\sigma_i^*}^2)
\end{align*}

where \(\phi(.)\) is the probability density function and \(\Phi(.)\) is
the cumulative distribution function of the standard normal
distribution.

\hypertarget{binary-missing-variable}{%
\paragraph{Binary missing variable}\label{binary-missing-variable}}

When \(y_{ij}^{miss}\) is a binary variable, the imputed value is drawn
from a Bernoulli distribution with a proportion parameter \(p_{ij}^*\)
given by \(P[y_{ij}=1|r_{ij}=0]\), that is the conditional probability
that \(y_{ij}=1\) given that the measure is unobservable (\(r_{ij}=0\)).
The \(p_{ij}^*\) is obtained from a bivariate probit model, as follows:
(Greene 2018) \begin{align*}
p^*_{ij} &= P[y_{ij}=1|r_{ij}=0,{\boldsymbol{\beta^O_{i}}^*},{\boldsymbol{\beta^{S}_{i}}^*},{\rho_{i}}^*]\\
p^*_{ij} &=\frac{\Phi_2(\boldsymbol{x^O_{ij}}\boldsymbol{{\beta^{O}_i}^*},-\boldsymbol{x^S_{ij}}\boldsymbol{{\beta^{S}_i}^*},-\rho_i)}{\Phi(-\boldsymbol{x^{S}_{ij}}{\boldsymbol{\beta^{S}_i}^*})}\\
y_{ij}^{miss}&\sim Ber(p^*_{ij})
\end{align*} where \(\Phi_2(.)\) corresponds to the bivariate normal
cumulative distribution function.

\hypertarget{multivariate-imputation}{%
\subsection{Multivariate imputation}\label{multivariate-imputation}}

When there are simultaneous missing variables in a dataset, our
imputation method can be extended in a Gibbs sampler procedure.
Particularly, our imputation method has been implemented according to
the structure of the MICE R package,(Buuren et al. 2021) that allows
imputing multiple incomplete predictors and covariates in a given
dataset.

Briefly, MICE (multiple imputation of chained equations) was built under
the fully conditional specification framework, where for each incomplete
variable a conditional imputation model is specified based on another
variables in the dataset. This process is carried out iteratively, so
that in each iteration the missing values of an incomplete variable are
drawn from the conditional distribution based on the updated variables
in the previous iteration.

Our imputation model can then be used in the imputation of any
incomplete variable in a dataset following an MNAR mechanism, even if it
is an outcome, predictor or auxiliary variable of the main analysis
model. Furthermore, as implemented according to the MICE framework, the
imputation method could be used simultaneously with other imputation
methods available in MICE or in add-in packages such as micemd (Audigier
and Resche-Rigon 2022), to impute datasets with multiple incomplete
variables that differ in type and missing mechanism.

\hypertarget{technical-details-of-the-implementation}{%
\subsection{Technical details of the
implementation}\label{technical-details-of-the-implementation}}

The Heckman model is estimated with the \textbf{gjrm} function of the
GJRM R package(Radice 2021), under the bivariate model with the
nonrandom sample selection (BSS) specification. The meta-analysis model
is performed with the \textbf{mixmeta} function of the R package
mixmeta(Gasparrini and Sera 2021), which allows the use of maximum
likelihood (ML), restricted maximum likelihood (REML), and moments
estimation methods. For the simulation and illustrative study, we use
the restricted REML estimation method, which is recommended as it has a
good balance between unbiasedness and efficiency. (Viechtbauer 2005).

\hypertarget{simulation-study}{%
\section{Simulation study}\label{simulation-study}}

\label{sec:simulation}

\hypertarget{aim}{%
\subsection{Aim}\label{aim}}

We designed a simulation study aimed to compare the performance of
alternative methods for imputing a single missing outcome variable in a
hierarchical dataset, where the missingness follows a MNAR mechanism. In
our scenarios we considered systematically missingness.

\hypertarget{data-generation-mechanism}{%
\subsection{Data-generation mechanism}\label{data-generation-mechanism}}

We generated the data from a Heckman selection model with bivariate
normal distribution error terms. For simplicity we started from ``basic
scenario'' where the database collected information from \(N=10\)
clusters of \(n_i=1000\) individuals. However, we altered both \(N\) and
\(n_i\) in sensitivity analyses (Table \ref{tab:scenarios}).

\begin{table}

\caption{\label{tab:scenarios}Data generation scenarios}
\centering
\begin{tabular}[t]{>{\raggedright\arraybackslash}p{2,5cm}>{\raggedright\arraybackslash}p{1,5cm}>{\raggedright\arraybackslash}p{2,9cm}>{\raggedright\arraybackslash}p{1,5cm}>{\raggedright\arraybackslash}p{2.7cm}}
\toprule
Scenario & Incomplete variable & $\rho$ & $N;n_i$ & Missing process\\
\midrule
\cellcolor{gray!6}{Base} & \cellcolor{gray!6}{Continuous} & \cellcolor{gray!6}{0.6} & \cellcolor{gray!6}{10;1000} & \cellcolor{gray!6}{Heckman (BVN)}\\
M(N)AR & Continuous, Binary & 0 (MAR), 0.3,0.6,0.9 (MNAR) & 10;1000 & Heckman (BVN)\\
\cellcolor{gray!6}{Size and cluster number} & \cellcolor{gray!6}{Continuous} & \cellcolor{gray!6}{0.6} & \cellcolor{gray!6}{10;50,   10;100,   10;1000,  50;1000,  100;1000} & \cellcolor{gray!6}{Heckman (BVN)}\\
Distribution deviations & Continuous & 0.6 & 10;1000 & Heckman (BVN), Heckman (t-skew), Explicit\\
\bottomrule
\end{tabular}
\end{table}

For each dataset, we generated \(X_{1i}\) a treatment indicator variable
from a Bernoulli distribution with a probability of treatment on each
cluster equal to 0.6. Next, we simulated the mean of two continuous
covariates from a multivariate normal distribution
\(\begin{psmallmatrix}\mu_2\\\mu_3\end{psmallmatrix}\sim N\left(\begin{psmallmatrix}0\\0\end{psmallmatrix},\begin{psmallmatrix}0.2 & 0.015\\0.015 & 0.2\end{psmallmatrix}\right)\).
We then simulated for each cluster a baseline covariate
\(X_{2i}\sim N(\mu_2,1)\) and a exclusion restriction variable
\(X_{3i} \sim N(\mu_3,0.5)\).

Here, we considered \(X_{1i}\) and \(X_{2i}\) as predictors in the
outcome equation, i.e.,
\(\boldsymbol{X_i}^O=[1,\boldsymbol{X_{1i}},\boldsymbol{X_{2i}}]\). For
the selection equation we included both variables and the \(X_{3i}\)
exclusion restriction variable,
\(\boldsymbol{X_i^S}=[1,\boldsymbol{X_{1i}},\boldsymbol{X_{2i}},\boldsymbol{X_{3i}}]\).
Then in case of a missing continuous variable, we calculate the latent
variables \(\boldsymbol{y_{i}^*}\) and \(\boldsymbol{r^*_{i}}\) as
follows: \begin{align*}
 \boldsymbol{y_{i}^*}&= \boldsymbol{\beta_i^O}\boldsymbol{X_i^O}+\boldsymbol{\epsilon^O_{i}} \\
 \boldsymbol{r_{i}^*}&= \boldsymbol{\beta_i^S}\boldsymbol{X_i^O}+\boldsymbol{\epsilon^S_{i}}
\end{align*} Here we assumed that all coefficient parameters varied
across studies, by including cluster-specific random effects as:
\begin{align*}
 \beta_{i}^O&= \beta^O + b_{i}^O\\
 \beta_{i}^S&= \beta^S + b_{i}^S
\end{align*} We fixed coefficients \(\beta^O=\{0.3,1,1\}\) and
\(\beta^S=\{-0.8,1.3,-0.7,1.2\}\) in order to get around 40\% of
sporadically missing values on the response \(y_{ij}\) in the entire
data set. Additionally, we ensured that the \(y_{ij}^*\) observations
were systematically missing in 20\% of the clusters included in the data
set, by removing the outcome values in the 20\% of the clusters.

We assumed that random effects were independent within equations
(\(b_{0}^O\rotatebox[origin=c]{90}{$\models$}b_{1}^O \rotatebox[origin=c]{90}{$\models$}b_{2}^O\)
and
\(b_{0}^S \rotatebox[origin=c]{90}{$\models$}b_{1}^S \rotatebox[origin=c]{90}{$\models$}b_{2}^S\rotatebox[origin=c]{90}{$\models$}b_{3}^S\)),
but were linked between both selection and outcome equations through a
bivariate normal distributed as: \[\begin{pmatrix}
b_i^O \\
b_i^S
\end{pmatrix}\sim N\left(\begin{pmatrix}
0 \\
0
\end{pmatrix},\psi_{kk}\begin{pmatrix}
1 & \rho*0.4 \\
\rho*0.4 &1 
\end{pmatrix}\right)
\] for the coefficients of intercept and the coefficients of the
variables \(X_1\),\(X_2\) with \(\psi_{00}=\psi_{11}=\psi_{22}=0.4\). We
considered that the correlation parameter of the random effects between
equations is 40\% of the value of the assumed correlation parameter
between error terms \(\rho\). In addition, we included a random effect
on the exclusion restriction variable given by \(b_3\sim N(0,0.2)\)
assuming that the intracluster variation in the exclusion restriction
effect is lower than the variation on other coefficient parameters
effects. The \(\rho\) parameter was given different values depending on
the simulated missing mechanism (Table \ref{tab:scenarios}) .

As regards the error terms, they were bivariate normal distributed as:
\[\begin{pmatrix}
\epsilon^O_{i} \\
\epsilon^S_{i}
\end{pmatrix}\sim N\left(\begin{pmatrix}
0 \\
0
\end{pmatrix},\begin{pmatrix}
\sigma^2_i & \rho\sigma_i \\
\rho\sigma_i & 1
\end{pmatrix}\right)
\] whose \(\sigma_i^2\) is a random parameter across clusters
distributed as \(log(\sigma_i) \sim N(0,0.05)\).

\hypertarget{adittional-scenarios}{%
\subsection{Adittional scenarios}\label{adittional-scenarios}}

In addition to the basic scenario described above, we explored
additional data generating mechanisms. Specifically we investigated the
performance of the imputation methods under the following scenarios:

\begin{itemize}
\item
  \textbf{M(N)AR scenarios:} We explored a missing MAR mechanism
  (\(\rho=0\)), a scenario when data followed a MNAR mechanism with a
  low (\(\rho=0.3\)), intermediate (\(\rho=0.6\)) and strong correlation
  (\(\rho=0.9\)) between \(y^*\) and \(r^*\). We also explored a
  scenario when the missing variable is binary. Therefore we simulated a
  \(y_{i}\) binary incomplete variable, by keeping similar parameters to
  the ones used in the simulation of a missing continuous variable, but
  here we defined the observable binary variable as: \begin{align*}
    r_{i}&=I(r*_{i}>0)\\
    y_{i}&=I(y^*_{i}>0) \forall r^*_{ij}>0
  \end{align*}
\item
  \textbf{Influence of sample size and cluster number}: we explored
  different configurations regarding the number of patients per cluster
  \(n_i=\{50,100,1000\}\) and the number of clusters
  \(N=\{10,50,100\}\).
\item
  \textbf{Violation of distributional assumptions:} Here, we aimed to
  investigate how the imputation models behave in settings with
  departures from the bivariate normal distribution, performed two
  sensitivity analyses.

  \begin{enumerate}[wide=10pt]
    \item \textbf{Skewed-t:} We drew error terms from a bivariate skewed student-t distribution using the same location parameter and covariance matrix of the normal distributed settings,  with 4 degrees of freedom and an $\alpha=\{-2,6\}$ parameter which regulates the asymmetry of the density.
    \item  \textbf{Explicit:} In addition we simulated an explicit missingness process, where error terms of the selection and outcome equations were independently normal distributed and the selection of observations depended on the value of the outcome variable, as $ry^*_i=0.3y^*_i+\epsilon^S_{i}$. These settings assure that the percentage of missingness on the outcome variable is around 60\% in all the evaluated scenarios.
    \item \textbf{Normal:} As a reference, we provide the results from the basic scenario of the main simulation study, where the error terms were Bivariate normal distributed. 
  \end{enumerate}
\end{itemize}

\hypertarget{estimands}{%
\subsection{Estimands}\label{estimands}}

The estimands were the parameter coefficients of the outcome equation
\(\beta^O=\{\beta^O_0,\beta^O_1,\beta^O_2\}\), with special emphasis on
the treatment effect parameter \(\beta^0_1\). We also report the
estimated standard deviation from the covariance matrix of the random
effects, i.e.,
\(\sqrt{\psi_{00}}\),\(\sqrt{\psi_{11}}\),\(\sqrt{\psi_{22}}\).

\hypertarget{estimating-procedures}{%
\paragraph{Estimating procedures:}\label{estimating-procedures}}

After the imputation procedure, we estimated the following (generalized)
mixed linear effect model using the \emph{lmer()} function from the
\emph{lme4} R package.(Bates et al. 2022)
\(y_{i}=\beta_i^OX_i^O+\epsilon^O_{i}\) In case of missing binary
variable, we used the same matrix of predictors but on a binary model
estimated with the \emph{glmer()} function from the \emph{lme4} R
package. Then, we pooled the estimates of the \(\beta_i^O\) and the
variance of the random effect and residual errors of the multiple
imputed datasets according to Rubin's rule(Rubin 1987), over which we
calculated the performance measures on the estimands.

To calculate the coverage of the parameter coefficients' 95\% confidence
intervals (CI), we estimate CI with the Wald method.

\hypertarget{imputation-methods}{%
\subsection{Imputation methods}\label{imputation-methods}}

For each scenario we simulated 500 datasets over which we evaluated the
following imputation methods:

\begin{itemize}
\item
  \textbf{Complete case analysis (CCA):} We removed all patients with
  missing observations.
\item
  \textbf{1l.Heckman:} Multiple imputation based on the Heckman model
  without no study specification, following the imputation method
  proposed by Galimard et al.(2016). (Galimard et al. 2016)
\item
  \textbf{2l.MAR}: Multiple imputation assuming MAR for hierarchical
  datasets, we used the multilevel imputation model ( 2l.2stage.norm and
  2l.2stage.bin) from the micemd R package(Audigier and Resche-Rigon
  2022), which are described in Audigier et al.~(2018) paper. (Audigier
  et al. 2018)
\item
  \textbf{2l.Heckman}: The proposed imputation method based on the
  Heckman model for hierarchical datasets.
\end{itemize}

\hypertarget{performance-measures}{%
\subsection{Performance measures}\label{performance-measures}}

We calculated the following measures, usually employed to evaluate
imputation methods(Buuren 2018), according to the formulas provided in
Morris et al.(2019)(Morris, White, and Crowther 2019):

\begin{itemize}
\tightlist
\item
  \textbf{Bias:} Bias on the coefficient and variance-covariance matrix
  terms.
\item
  \textbf{Coverage:} Coverage of the 95\% confidence intervals for the
  coefficient parameters.
\item
  \textbf{Width}: Average width of the confidence interval on the
  coefficient parameters.
\item
  \textbf{RMSE:} Root mean squared error of the coefficient and random
  effect parameters.
\end{itemize}

In addition, in the appendix table we reported the empirical standard
errors (EmpSE), Monte Carlo standard errors (ModSE) on the coefficient
parameters, average processing time (time in seconds) and the percentage
of datasets where the imputation method converged (run), i.e., the
imputation method generated an output.

\hypertarget{software}{%
\subsection{Software}\label{software}}

For the simulation study and illustrative examples we used R version
4.0.4 in a linux environment. (R Core Team 2021)

The Heckman 2L imputation method is available in the micemd R package
(Audigier and Resche-Rigon 2022) (as \textbf{mice.2l.2stage.heckman()})
and also on the github repository
\url{https://github.com/johamunoz/Statsmed_Heckman} where you can also
find all the codes accompanying this paper and a toy example that
explains how to implement the method in mice.

\hypertarget{results-from-the-simulation-study}{%
\subsection{Results from the simulation
study}\label{results-from-the-simulation-study}}

\hypertarget{results-mnar-scenarios}{%
\subsubsection{Results M(N)AR scenarios}\label{results-mnar-scenarios}}

\hypertarget{continuous-incomplete-variable}{%
\paragraph{Continuous incomplete
variable}\label{continuous-incomplete-variable}}

\begin{figure}[!htb]
 {\centering \includegraphics{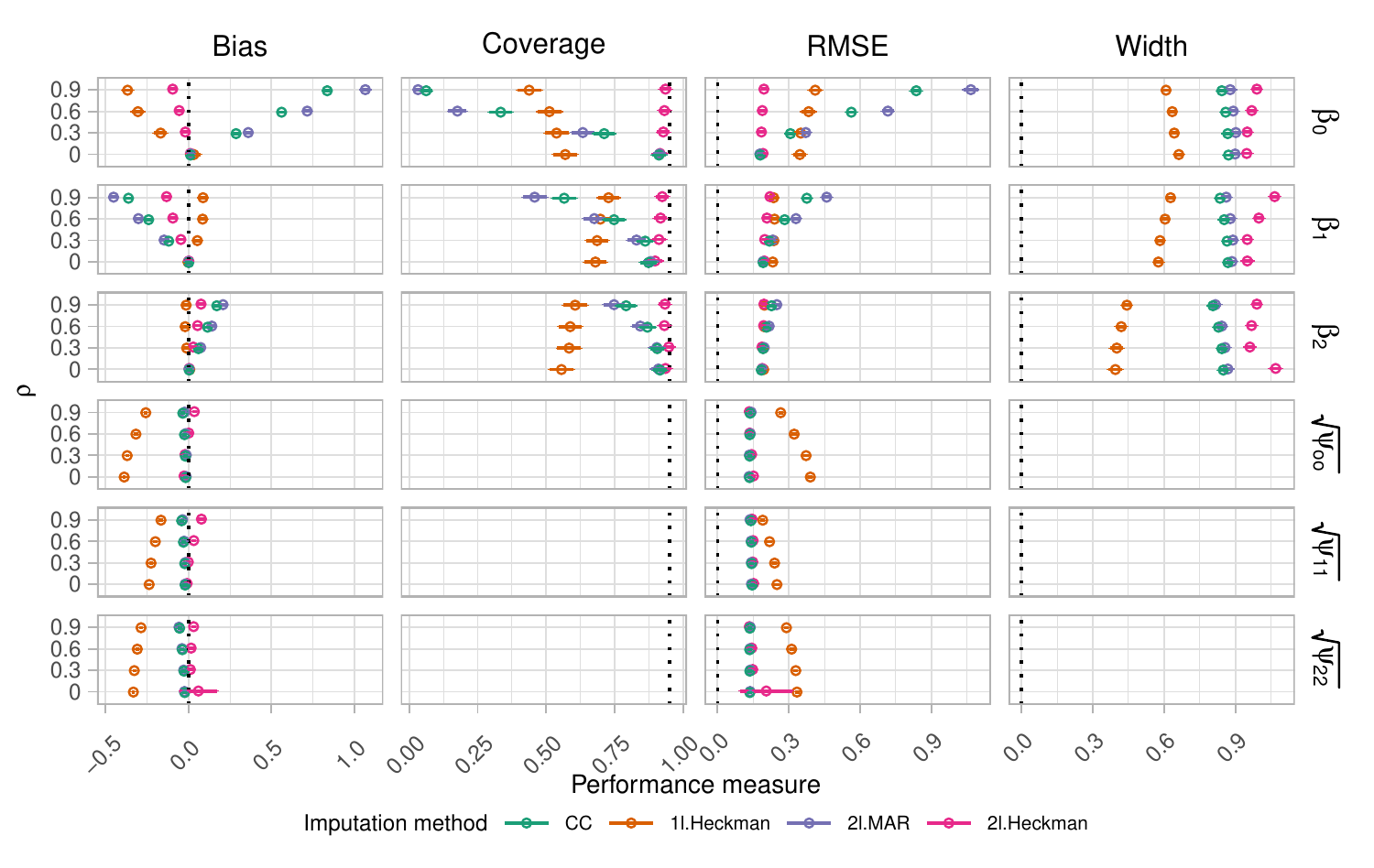} 
 }
 
 \caption{Continuous incomplete variable by varing $\rho$, where dashed lines depict the target performance criteria value}\label{fig:rhocont}
 \end{figure}

Figure \ref{fig:rhocont} shows the results of simulations where the
missing variable was continuous. In the MAR scenario, i.e.~when
\(\rho=0\), all imputation methods provided similar unbiased estimates
of the coefficient parameters, but as rho increased, i.e.~the mechanism
became MNAR, the estimates for the complete case analysis and the
MAR-IPD imputation method became biased.

\begin{figure}[!htb]

{\centering \includegraphics{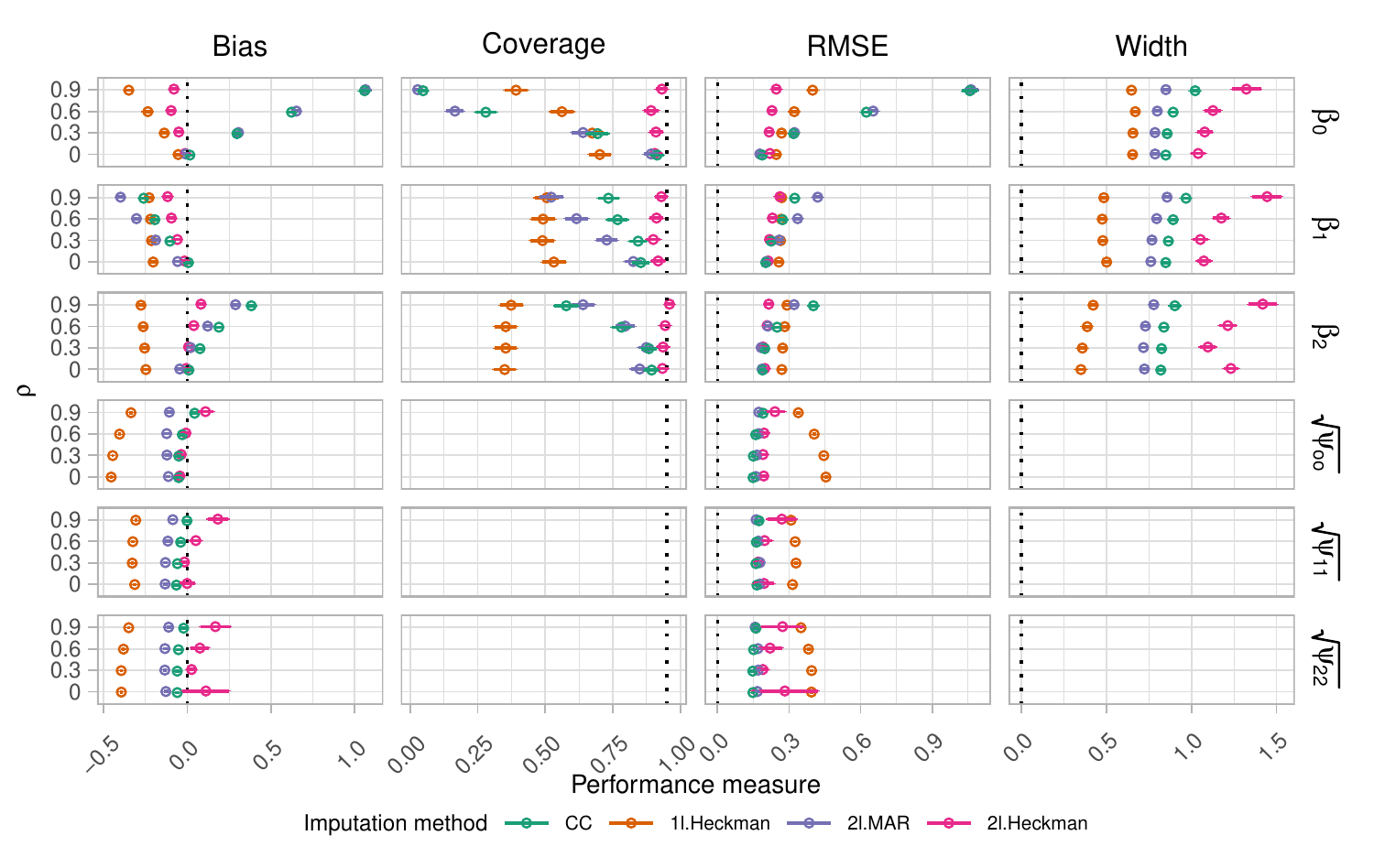} 

}

\caption{Binary incomplete variable by varing $\rho$, where dashed lines depict the target performance criteria value}\label{fig:rhobin}
\end{figure}

As it is expected both Heckman-based imputation (1l. and 2l.) models
provided less biased estimates of the coefficient parameters in MNAR
scenarios, but the random effects estimates on the 1l.heckman were far
away from the true values as no cluster information was considered at
all.

Regarding coverage, the 2l.Heckman imputation method provided the best
coverage values across all the \(\rho\) scenarios with a coverage level
close to the nominal 95\% interval. Even though 2l.Heckman was not
properly a randomization-valid method (Buuren 2018), as it was not
unbiased and had a coverage above 95\% across all the \(\rho\) values,
it led to better results in terms of bias and coverage compared to the
evaluated methods.

The 2l.Heckman method, also resulted in better estimates in terms of
RMSE than the other methods evaluated, via a better trade-off between
bias and variance.

In particular, our method provided an advantage over the 1l.Heckman
method in systematic missingness scenarios, by adding a cluster variable
in the imputation model, a feature lacking from 1l.Heckman. This can be
seen in the estimates of the random effects parameters of the 1l.Heckman
method, which were more biased than those of the other methods in which
cluster information was included, i.e.~2l.MAR and 2l.Heckman.

Regarding the width of the 95\% CI of the estimated parameters, we
observed that using the 2l.Heckman model we obtained larger CIs than
when using other evaluated methods, which generally increased as
\(\rho\) moved away from zero. In particular, we observed in \(\beta_2\)
that the CI length was larger in the MAR scenario. By checking the
simulation results further, we noted that the high variability came from
one simulation in which the linear model estimation had singularity
problems (not shown here).

\hypertarget{bivariate-incomplete-variable}{%
\paragraph{Bivariate incomplete
variable}\label{bivariate-incomplete-variable}}

In the case of a missing bivariate variable, we also found (Figure
\ref{fig:rhobin}) that the 2l.Heckman method provided the least unbiased
results on the coefficient parameters. However, we observed more
unbiased estimates and a larger 95th percentile amplitude in the
estimates of the standard deviations of the random effects, especially
in the \(\sqrt{\psi_{22}}\) compared to the estimates obtained in the
continuous incomplete outcome scenario.

\hypertarget{sensitivity-analysis-number-of-clusters-and-sample-size-of-clusters}{%
\subsubsection{Sensitivity analysis: number of clusters and sample size
of
clusters}\label{sensitivity-analysis-number-of-clusters-and-sample-size-of-clusters}}

\begin{figure}[!htb]

{\centering \includegraphics{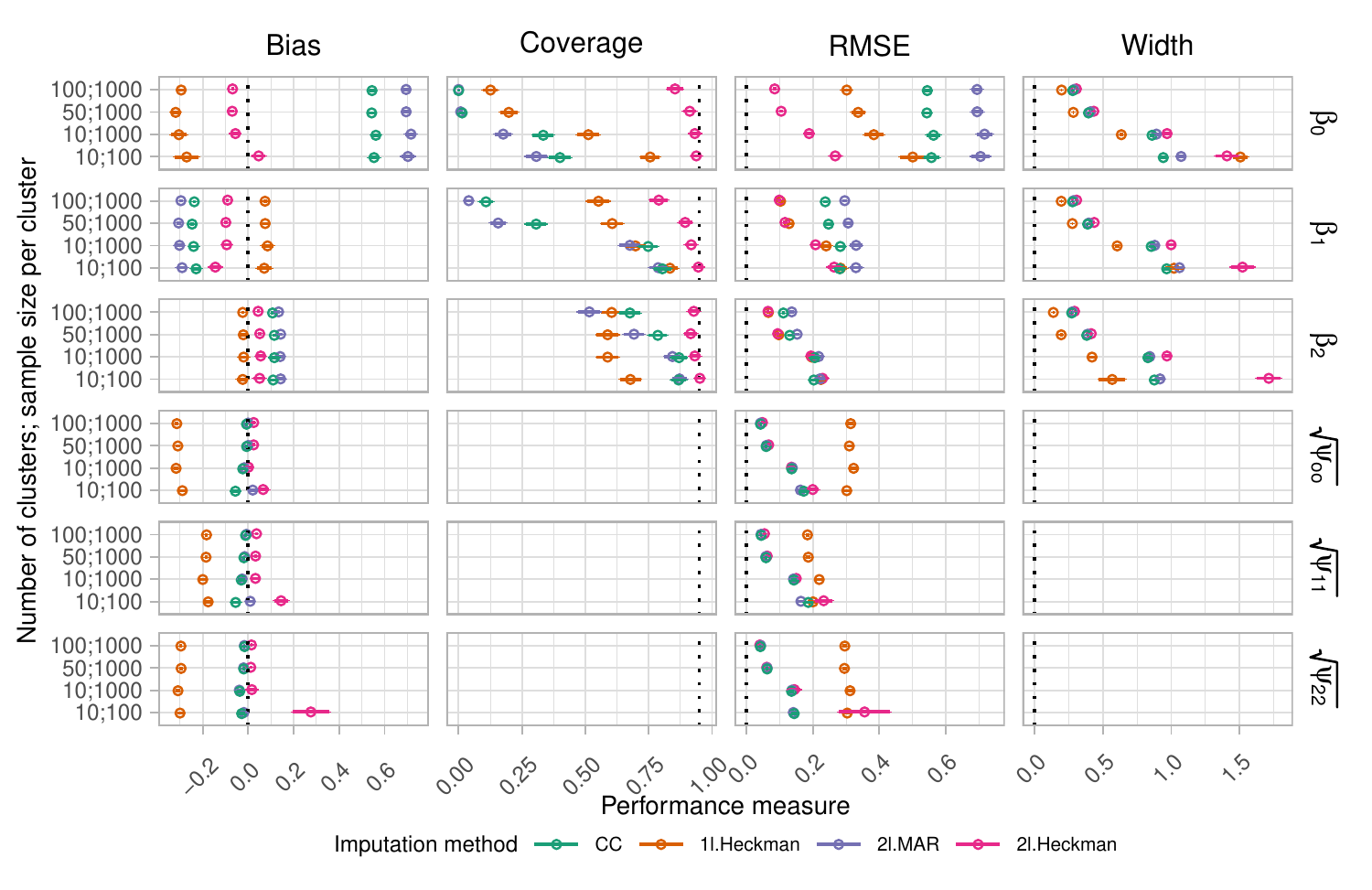} 

}

\caption{Continuous incomplete variable under systematical missingness by varing: number of clusters (N); sample size per cluster ($n_i$), where dashed lines depict the target performance criteria value}\label{fig:size}
\end{figure}

We evaluated how robust our method was to variations in the number of
clusters and also in the sample size of cluster (Figure \ref{fig:size}).

By increasing N, the number of clusters, from 10 to 100, we observed
that the bias was not affected but the width of the 95\% CI of the
estimates decreased (larger precision) and hence the RMSE decreased. On
the other hand, by reducing the number of units per cluster (from
\(n_i\)=1000 to \(n_i\)=100) the precision decreased for all
coefficients, the bias on coefficient estimates were not drastically
affected but the bias of random effects did.

When we reduced the sample size to 50 patients per study, the bias and
RMSE of the \(\sqrt{\psi_{22}}\) were drastically affected (not shown
here but in the Appendix). This could be explained in part due to the
scarce information on certain clusters, which affects directly the
estimation of the Heckman model on those clusters.

\hypertarget{sensitivity-analysis-distributional-assumptions}{%
\subsubsection{Sensitivity analysis: distributional
assumptions}\label{sensitivity-analysis-distributional-assumptions}}

When using the Heckman model in an explicit MNAR process, i.e.~when the
probability of loss is associated with the value of the missing
variable, we observe that under all imputation methods we obtain biased
estimators, with poor coverage. Particularly when estimating the
intercept we observed more biased estimates compared to those obtained
for the other estimated parameters \(\beta_2\) with coverage below 60\%.

Our model may not be fully suitable for this type of scenario,
especially if the main analysis is focused on estimating absolute
prevalence estimates, as it is greatly affected by the bias of the
intercept parameter (\(\beta_0\)) and the width of the 95th percentile
CI of all coefficient parameters. Also the imputation method was
affected in terms of the bias of the standard deviation of the
random-effects parameters \(\sqrt{\psi_{22}}\).

With respect to the Skewed-t scenario, only the bias of \(\beta_0\) was
affected for all imputation methods applied. But \(\beta_0\) estimates
for both the 1l.Heckman and 2l.Heckman were drastically affected in
terms of bias, with very poor coverage (below 25\%).Therefore, the
applicability of our method could be questionable in scenarios that do
not conform to the BNV assumption.

\begin{figure}[!htb]

{\centering \includegraphics{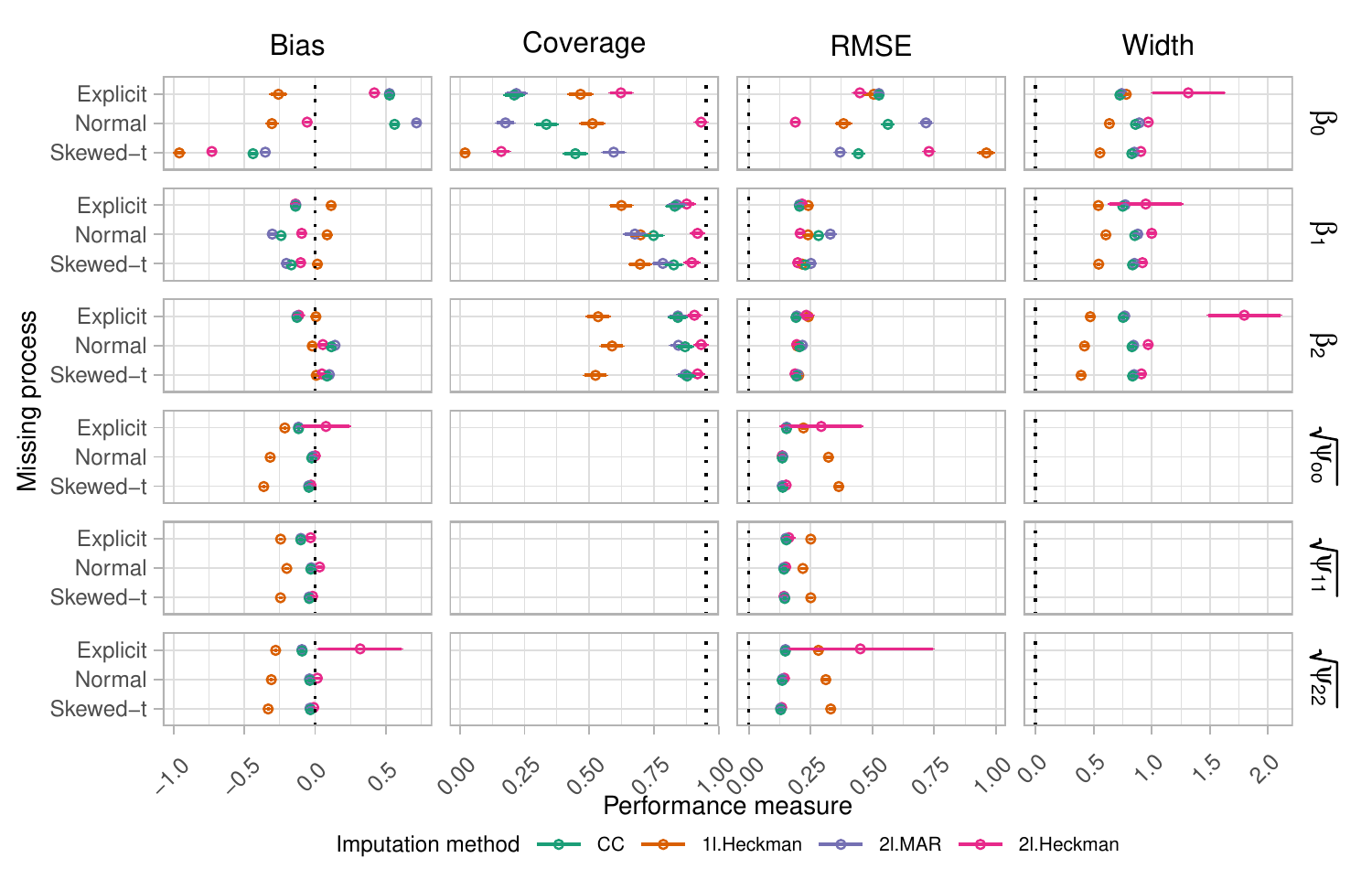} 

}

\caption{Continuous incomplete variable with deviations in distribution assumptions, where dashed lines depict the target performance criteria value}\label{fig:dist}
\end{figure}

\hypertarget{an-illustrative-study}{%
\section{An illustrative study}\label{an-illustrative-study}}

\label{sec:illustration}

Malaria is a mosquito-borne disease and is the leading cause of illness
and death in Africa, especially in children and pregnant women. To
prevent the spread of the disease, long-lasting nets (LLINs) and indoor
residual spraying (IRS) in at-risk households are used as control
measures.

Specifically, in Uganda, under the Uganda LLIN evaluation project, a
LLINS distribution campaign was conducted between 2013 and 2014. In
2017, the effect of LLIN control together with insecticides was assessed
through a cross-sectional community survey in 104 health sub-districts
in 48 districts located within 5 sub-regions of Uganda.

In each sub-district, a sample of households with at least one child
aged 2-10 years was surveyed, where information was collected on
household conditions and use of preventive measures. In addition, finger
prick blood samples were taken from each child to determine the
prevalence of parasitemia and an entomological study was conducted to
estimate mosquito prevalence. Details of the project and survey are
provided elsewhere. (S. G. Staedke et al. 2019)

For this example, we used data accessed directly from ClinEpiDB(S.
Staedke et al. 2021), where data were collected from 5195 households
with verified consent, inhabited by 11137 residents aged 2-10 years.
Blood samples were only taken from 8846 children, as 69 were excluded
from the study due to lack of consent and 2222 were not present at the
time of the survey. Although the original data set consists of 164
variables, here we only consider the variables described in Table
\ref{tab:descriptive}, which were used as predictors in the imputation
model and are fully observed in the dataset. In this dataset, the
parasitaemia test is an incomplete binary result (1=positive test,
0=negative test), which is missing 21\% across the whole dataset.

\begin{table}

\caption{\label{tab:descriptive}Descriptive analysis, predictor variables}
\centering
\resizebox{\linewidth}{!}{
\begin{tabular}[t]{>{\raggedleft\arraybackslash}p{2,2cm}>{\raggedleft\arraybackslash}p{1cm}>{\raggedleft\arraybackslash}p{1cm}>{\raggedleft\arraybackslash}p{1,5cm}>{\raggedleft\arraybackslash}p{2,2cm}>{\raggedleft\arraybackslash}p{2,2cm}>{\raggedleft\arraybackslash}p{1cm}>{\raggedleft\arraybackslash}p{1cm}>{\raggedleft\arraybackslash}p{1cm}>{\raggedleft\arraybackslash}p{1cm}}
\toprule
Sub-region & District (N) & Children (N) & Age mean (years) & Log10 Female Anopheline & Wealth index & Bednet (\%) & Girls (\%) & Holiday (\%) & No test (\%)\\
\midrule
\cellcolor{gray!6}{North East} & \cellcolor{gray!6}{5} & \cellcolor{gray!6}{794} & \cellcolor{gray!6}{5.50} & \cellcolor{gray!6}{2.67[1.5,4.3]} & \cellcolor{gray!6}{-0.45[-1.2,2.2]} & \cellcolor{gray!6}{10.7} & \cellcolor{gray!6}{49.0} & \cellcolor{gray!6}{31.9} & \cellcolor{gray!6}{17.5}\\
Mid Eastern & 8 & 1354 & 5.61 & 0.84[0.1,2.5] & -0.14[-1.0,2.5] & 9.3 & 48.1 & 32.9 & 25.6\\
\cellcolor{gray!6}{South Western} & \cellcolor{gray!6}{14} & \cellcolor{gray!6}{3596} & \cellcolor{gray!6}{5.69} & \cellcolor{gray!6}{0.27[0.1,1.3]} & \cellcolor{gray!6}{0.18[-1.0,2.9]} & \cellcolor{gray!6}{23.8} & \cellcolor{gray!6}{49.4} & \cellcolor{gray!6}{66.5} & \cellcolor{gray!6}{21.1}\\
Mid Western & 12 & 3172 & 5.66 & 1.27[0.1,3.2] & -0.03[-1.0,2.8] & 13.3 & 48.9 & 62.9 & 20.5\\
\cellcolor{gray!6}{East Central} & \cellcolor{gray!6}{9} & \cellcolor{gray!6}{2152} & \cellcolor{gray!6}{5.61} & \cellcolor{gray!6}{2.74[0.4,6.3]} & \cellcolor{gray!6}{0.01[-1.1,3.1]} & \cellcolor{gray!6}{13.2} & \cellcolor{gray!6}{51.6} & \cellcolor{gray!6}{51.6} & \cellcolor{gray!6}{16.0}\\
\bottomrule
\end{tabular}}
\end{table}

To illustrate our proposed method, following the article by Rugnao et
al.~(2019),(Rugnao et al. 2019) we estimated the prevalence of
parasitemia by subregion and by age after approximately 3 years of LLIN
campaigns started. We estimated parasitemia prevalence using 3
approaches that made different assumptions on the missingness mechanism:
MCAR, MAR and MNAR.

Under the MCAR assumption, prevalence was calculated on the basis of the
recorded tests, i.e., we only included patients with a test result.
Under the MAR assumption, the test values of children who were not
present during the survey were imputed with the 2l.2stage.bin method of
the micemd package, where the community was taken as the cluster and the
following factors previously associated with parasitemia were used as
predictors in the imputation model: sex, bednet ( indicator of whether
only two or fewer persons share a mosquito bed net). In addition, we
included age as a power 3 spline function, the cluster-level Log10 mean
of the number of female anopheline mosquitoes per household estimated
from the entomological survey, and the household wealth index from
principal components analysis calculated specifically for the surveyed
households.

Under the MNAR assumption, we used the proposed 2l.Heckman method to
impute missing test values. The selection and outcome equation included
the same predictor variables as used under the MAR approach. In
addition, we included a holiday indicator variable as ERV. This was
calculated according to school vacation calendars and public holidays in
Uganda in 2017. We examined the association of this ERV with the outcome
variable (y) and with the selection indicator (ry), conditioned on the
remaining imputation predictors. The model results in Table
\ref{tab:exclusion} indicate that the holiday indicator could be a
plausible ERV variable, as there was strong evidence of an association
with ry, but no evidence of an association with y.

\begin{table}

\caption{\label{tab:exclusion}Evaluation of Holidays as exclusion restriction variable}
\centering
\begin{tabular}[t]{>{\raggedright\arraybackslash}p{2,5cm}>{\raggedright\arraybackslash}p{2,5cm}>{\raggedright\arraybackslash}p{2,5cm}}
\toprule
Predictors/response & Test result (y) & Test taken (ry)\\
\midrule
\cellcolor{gray!6}{(Intercept)} & \cellcolor{gray!6}{-1.07(0.05)***} & \cellcolor{gray!6}{1.35(0.05)***}\\
Log10 Female Anopheline & 0.73(0.03)*** & 0.15(0.02)***\\
\cellcolor{gray!6}{Wealth index} & \cellcolor{gray!6}{-0.55(0.04)***} & \cellcolor{gray!6}{0.04(0.03)}\\
Bednet-Yes & -0.30(0.08)*** & 0.70(0.08)***\\
\cellcolor{gray!6}{Holidays-Yes} & \cellcolor{gray!6}{-0.04(0.05)} & \cellcolor{gray!6}{0.19(0.05)***}\\
\addlinespace
Girls-No & 0.10(0.05) & 0.05(0.05)\\
\cellcolor{gray!6}{s(Age)} & \cellcolor{gray!6}{1.84(1.97)***} & \cellcolor{gray!6}{1.02(1.04)***}\\
\bottomrule
\multicolumn{3}{l}{\rule{0pt}{1em}\textit{Note: } ***p<0.01}\\
\end{tabular}
\end{table}

\begin{figure}[!htb]

{\centering \includegraphics{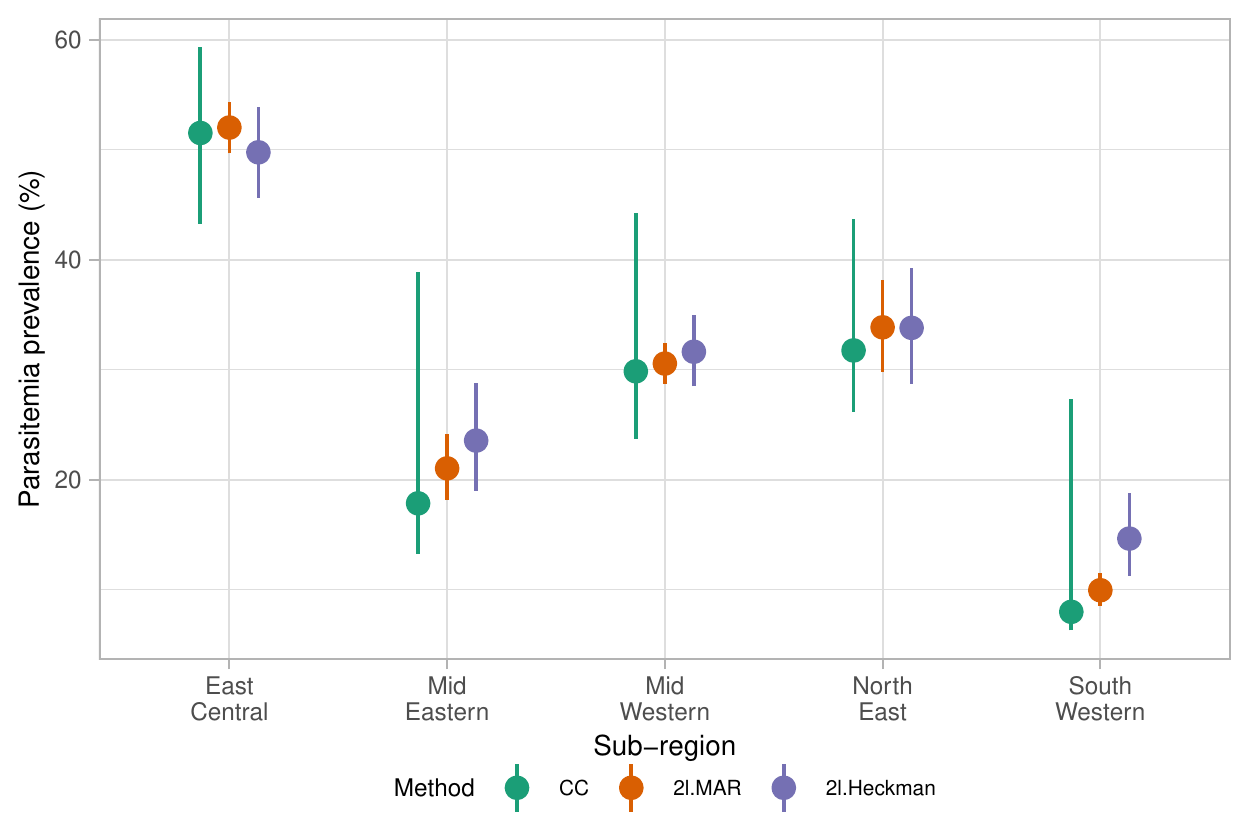} 

}

\caption{Estimates of malaria prevalence by sub-region}\label{fig:region}
\end{figure}
\begin{figure}[!htb]

{\centering \includegraphics{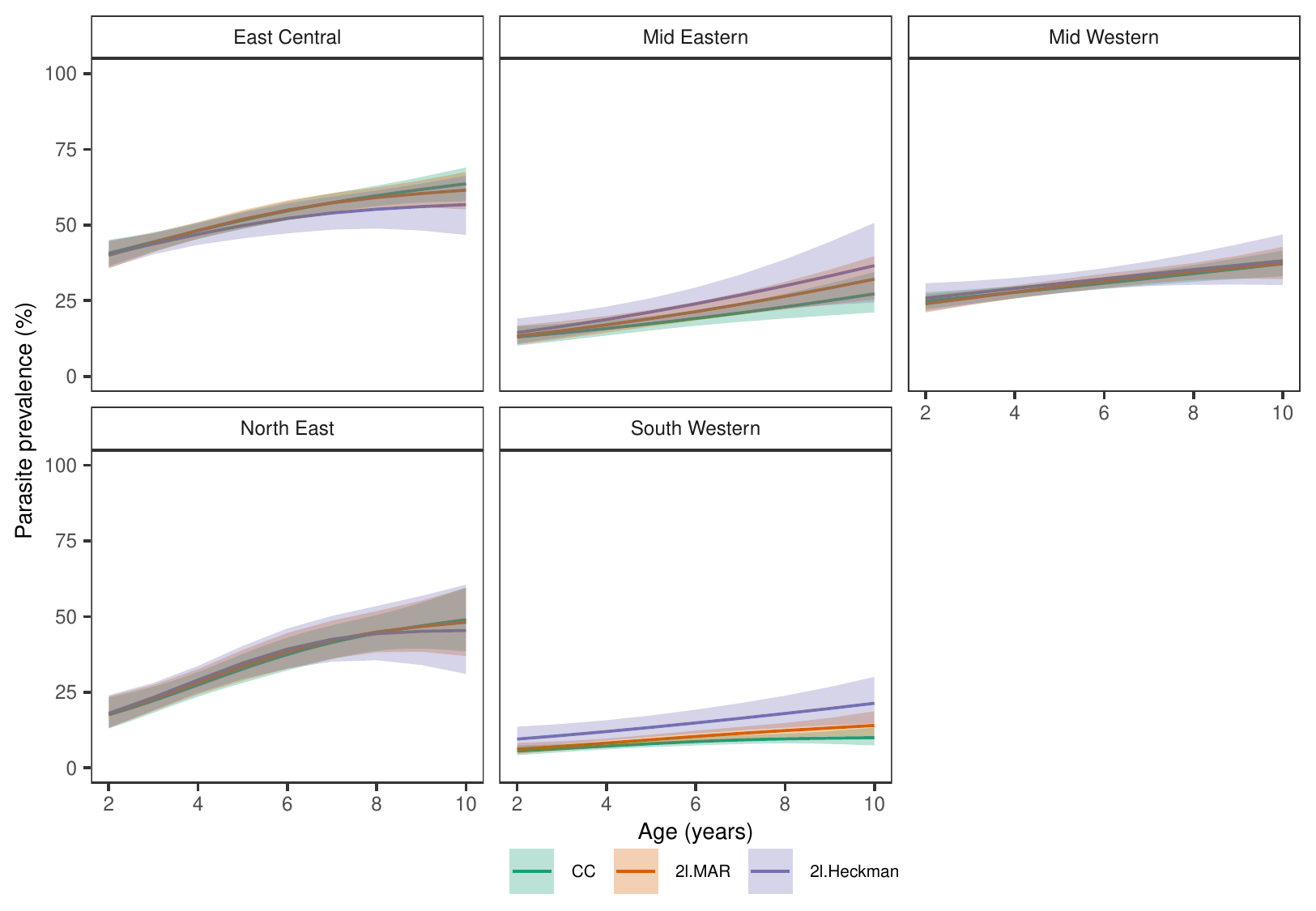} 

}

\caption{Estimates of malaria prevalence by sub-region and age}\label{fig:region_age}
\end{figure}

According to our imputation approach, non-tested children were estimated
to have a higher prevalence of malaria than participants in more than
half of the districts analyzed. As can be seen in Figure
\ref{fig:region}, for each subregion the prevalence estimates of the
approaches did not differ significantly between methods. However,
prevalence estimates under the MNAR assumption (i.e.~2 level Heckman)
were higher than those estimated under the MAR or MCAR aproaches, except
for the East-Central region.

In terms of prevalence by age, there were no significant differences
between methods (Figure \ref{fig:region_age}). The prevalence estimates
for children aged 2 to 6 years were very similar in all regions under
the different assumptions. Assuming that children start going to school
after the age of 6, the results could be partly explained by the
mobility of 2-6 year olds compared to school-age children,
i.e.~school-age children spend more time outdoors and travel more than
younger children.

However for school children, prevalences estimated with the Heckman
method were found to be higher in the Mid-East and Southwest regions
than those obtained with the other methods, whereas in the East-Central
region the estimates with the Heckman method are lower. A possible
reason for selection bias in surveys of this type is, for example, that
daytime visits might favor measurement in sick school children who stay
home, leading to overestimated prevalence results as found in the
East-Central region. (Program 2020) Nevertheless, we were unable to find
information confirming the direction in which malaria prevalence is
driven by selection bias in this Uganda study or in other studies
similar to this one.

\hypertarget{discussion}{%
\section{Discussion}\label{discussion}}

\label{sec:discussion} We have extended and evaluated methods for
multiple imputation of clustered datasets, in situations where some
incomplete variables follow a MNAR mechanism. For clustered datasets,
only imputation methods under the MAR mechanism had previously been
proposed. Although there are imputation methods that can handle MNAR
they can only handle the case of individual studies. This puts limit to
their use in situations common in IPD-MA such as when there is
systematic missingness or when the proportion of missingness of a
variable is very high in one of the included studies. To address this
gap, we proposed a new multiple imputation method for continuous and
binary MNAR covariates, that is specifically designed for clustered
datasets. Our method, allows borrowing of information between the
clusters to obtain more reliable imputation results at the individual
cluster level.

In our simulations we observed that the imputation method we proposed is
optimal for the imputation of continuous and binary missing variables
that follow a MNAR mechanism according to the Heckman model and that
come from multilevel data, such data commonly used in the IPDMA.

Overall, our new method produced unbiased estimates with convergence
close to 95\% for the coefficient parameters. It also resulted in less
biased estimates for the random effects parameters compared to the other
methods evaluated.

Empirically, we showed that the proposed method could be applicable when
there is systematic missigness at the cluster level. This method, in
particular, could provide less biased imputation values compared with
individual-level imputation methods, as it not only allows imputation of
missing values in clusters with systematic missigness, but can also
shrunk the values of individual clusters toward the overall study mean.
This is especially advantageous in studies with small sample sizes,
where an analysis approach that ignores data from other studies may lead
to extreme estimates of prevalence. By using our proposed method, these
would be reduced to the overall mean.

The advantage of the proposed method over methods that assume MAR is
that it allows the imputation of variables from cluster level data
following a MAR or MNAR mechanism according to the Heckman model. That
is to say that under the specification of a valid exclusion variable the
method determines which is the most adjustable correlation parameter
between equations (\(\rho\)), or in general terms the missingness
mechanism (MAR or MNAR), in each of the clusters evaluated.

Our implementation of the imputation method was built according to the
specifications of the mice R package and is available in the micemd
package, which allows, first of all, to be used both on the outcome and
on the covariates. In addition, it offers the option of being used
simultaneously with other imputation methods implemented in the MICE
package, which is advantageous in databases containing missing variables
with different prediction methods and models. Finally, the method can be
used on systematically and sporadically missing clusters, both for
continuous and binary missing variables with heterogeneous effects and
error variances.

\hypertarget{limitations-and-future-directions}{%
\subsection{Limitations and future
directions}\label{limitations-and-future-directions}}

A major limitation of our method is that it needs a valid restriction
variable, which in some contexts is difficult to establish at the
individual study level and can be even more challenging if one tries to
find a valid exclusion variable across clusters.

In addition, to estimate the marginal estimates, the method only uses
clusters with observable information, i.e.~that are not systematically
missing or have sufficient information to estimate the Heckman model.
The latter might restrict the evaluation of the Heckman model at the
cluster level to a certain number of predictors depending on the sample
size of the cluster.

Also, the method can be sensitive to both the sample size and the number
of studies included in the database. On the one hand, a small sample
size at the individual study level can affect not only the precision of
estimates but also the convergence of the method since sample size
needed to estimate the parameters of the Heckman model which can be at
least twice the number of parameters required to estimate in an
imputation model that assumes MAR. On the other hand, a high number of
studies that may improve the precision of the estimators may also make
the estimation of the marginal parameters more difficult and also
considerably increase the processing time of our method.

In our simulation study, data were generated by assuming a constant
correlation across all clusters in order to evaluate the performance
against M(N)AR assumptions. In practice, however this parameter can be
variable across clusters and this can considerably affect the
performance of our method. Therefore, in future research the effect of
this parameter could be further evaluated. One might also consider
relaxing this assumption of constant correlation to allow for a random
effects distribution for the correlation parameter.

Further, the method can also be extended to other copula models for
non-random selection, with different distributions of the selection and
outcome equations and dependency structure.(Wojtyś, Marra, and Radice
2018) Similarly, less restrictive Heckman based models can be considered
in terms of normality distribution of errors and no specification of
exclusion variables such as those proposed by Ogundimu \& Collins
(2019).(Ogundimu and Collins 2019).

\hypertarget{conclusion}{%
\subsection{Conclusion}\label{conclusion}}

We have proposed an extension to the Heckman model that can account for
MNAR, MAR or MCAR of a continuous or binary variable in clustered data
sets. Our simulations showed that it can have favorable statistical
properties, when its assumptions were met, and provided that the sample
size is sufficiently large. Regarding deviations from distributional
assumptions of the error terms, the coefficient parameters were fairly
robust in terms of bias, but the intercept was not.

\hypertarget{footnotes}{%
\section*{Footnotes}\label{footnotes}}
\addcontentsline{toc}{section}{Footnotes}

\hypertarget{disclaimer}{%
\subsubsection*{Disclaimer}\label{disclaimer}}

The views expressed in this paper are the personal views of the authors
and may not be understood or quoted as being made on behalf of or
reflecting the position of the regulatory agency/agencies or
organizations with which the authors are employed/affiliated.

\hypertarget{acknowledgements}{%
\subsubsection*{Acknowledgements}\label{acknowledgements}}

\begin{minipage}{.08\textwidth}
\includegraphics[width=\textwidth]{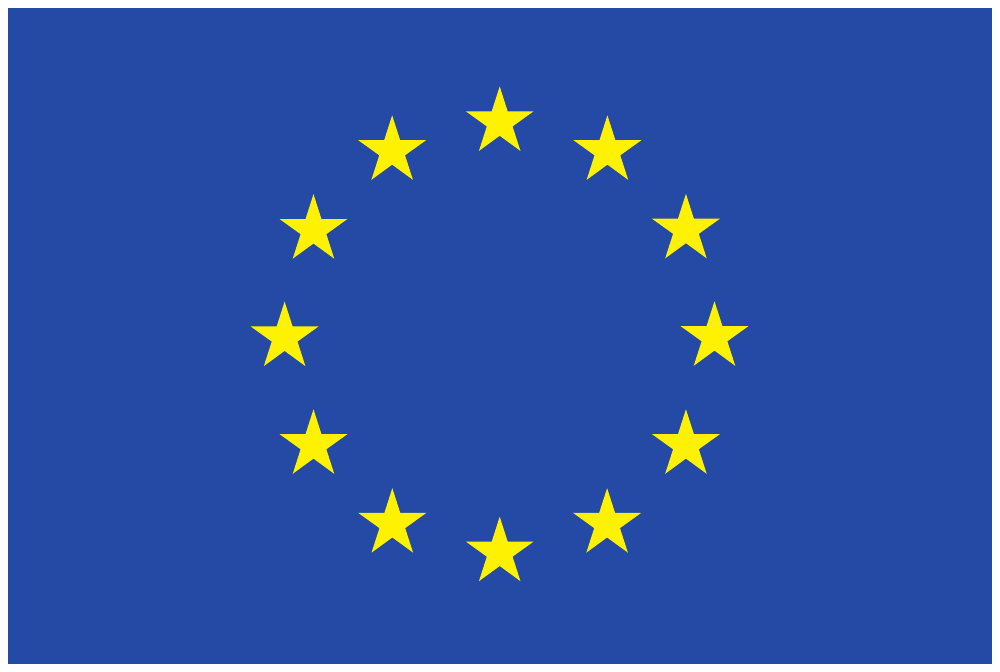}
\end{minipage}
\begin{minipage}{.90\textwidth}
This project has received funding from the European Union's Horizon 2020 research and innovation programme under ReCoDID grant agreement No 825746.
\end{minipage}

\hypertarget{References}{%
\section*{References}\label{references}}

\begin{CSLReferences}{1}{0}
\leavevmode\vadjust pre{\hypertarget{ref-amemiya1984}{}}%
Amemiya, Takeshi. 1984. {``Tobit Models: {A} Survey.''} \emph{Journal of
Econometrics} 24 (1): 3--61.
\url{https://doi.org/10.1016/0304-4076(84)90074-5}.

\leavevmode\vadjust pre{\hypertarget{ref-angrist2001}{}}%
Angrist, Joshua D., and Alan B. Krueger. 2001. {``Instrumental
{Variables} and the {Search} for {Identification}: {From Supply} and
{Demand} to {Natural Experiments}.''} \emph{Journal of Economic
Perspectives} 15 (4): 69--85. \url{https://doi.org/10.1257/jep.15.4.69}.

\leavevmode\vadjust pre{\hypertarget{ref-audigier2021}{}}%
Audigier, Vincent, and Matthieu Resche-Rigon. 2022. {``Micemd: Multiple
Imputation by Chained Equations with Multilevel Data.''}
https://CRAN.R-project.org/package=micemd.
\href{https://R\%20package\%20version\%201.9.0}{R package version
1.9.0}.

\leavevmode\vadjust pre{\hypertarget{ref-audigier_etal18}{}}%
Audigier, Vincent, Ian R. White, Shahab Jolani, Thomas P. A. Debray,
Matteo Quartagno, James Carpenter, Stef van Buuren, and Matthieu
Resche-Rigon. 2018. {``Multiple {Imputation} for {Multilevel Data} with
{Continuous} and {Binary Variables}.''} \emph{Statistical Science} 33
(2). \url{https://doi.org/10.1214/18-STS646}.

\leavevmode\vadjust pre{\hypertarget{ref-bates2022}{}}%
Bates, Douglas, Martin Maechler, Ben Bolker {[}aut, cre, Steven Walker,
Rune Haubo Bojesen Christensen, Henrik Singmann, et al. 2022. {``Lme4:
{Linear Mixed-Effects Models} Using '{Eigen}' and {S4}.''}
https://CRAN.R-project.org/package=lme4.

\leavevmode\vadjust pre{\hypertarget{ref-buuren18c}{}}%
Buuren, Stef van. 2018. \emph{Flexible Imputation of Missing Data}. 2nd
edition. Chapman and {Hall}/{CRC} Interdisciplinary Statistics Series.
{Boca Raton}: {CRC Press, Taylor and Francis Group}.

\leavevmode\vadjust pre{\hypertarget{ref-buuren2021}{}}%
Buuren, Stef van, Karin Groothuis-Oudshoorn, Gerko Vink, Rianne
Schouten, Alexander Robitzsch, Patrick Rockenschaub, Lisa Doove, et al.
2021. {``Mice: {Multivariate Imputation} by {Chained Equations}.''}
https://CRAN.R-project.org/package=mice.

\leavevmode\vadjust pre{\hypertarget{ref-galimard_etal18}{}}%
Galimard, Jacques-Emmanuel, Sylvie Chevret, Emmanuel Curis, and Matthieu
Resche-Rigon. 2018. {``Heckman Imputation Models for Binary or
Continuous {MNAR} Outcomes and {MAR} Predictors.''} \emph{BMC Medical
Research Methodology} 18 (1): 90.
\url{https://doi.org/10.1186/s12874-018-0547-1}.

\leavevmode\vadjust pre{\hypertarget{ref-galimard_etal16}{}}%
Galimard, Jacques-Emmanuel, Sylvie Chevret, Camelia Protopopescu, and
Matthieu Resche-Rigon. 2016. {``A Multiple Imputation Approach for
{MNAR} Mechanisms Compatible with {Heckman}'s Model.''} \emph{Statistics
in Medicine} 35 (17): 2907--20. \url{https://doi.org/10.1002/sim.6902}.

\leavevmode\vadjust pre{\hypertarget{ref-gasparrini2021}{}}%
Gasparrini, Antonio, and Francesco Sera. 2021. {``Mixmeta: {An Extended
Mixed-Effects Framework} for {Meta-Analysis}.''}
https://CRAN.R-project.org/package=Mixmeta.

\leavevmode\vadjust pre{\hypertarget{ref-gomes_etal20}{}}%
Gomes, Manuel, Michael G. Kenward, Richard Grieve, and James Carpenter.
2020. {``Estimating Treatment Effects Under Untestable Assumptions with
Nonignorable Missing Data.''} \emph{Statistics in Medicine} 39 (11):
1658--74. \url{https://doi.org/10.1002/sim.8504}.

\leavevmode\vadjust pre{\hypertarget{ref-greene2018}{}}%
Greene, William H. 2018. \emph{Econometric Analysis}. Eighth edition.
{New York, NY}: {Pearson}.

\leavevmode\vadjust pre{\hypertarget{ref-hammon_zinn20}{}}%
Hammon, Angelina, and Sabine Zinn. 2020. {``Multiple Imputation of
Binary Multilevel Missing Not at Random Data.''} \emph{Journal of the
Royal Statistical Society: Series C (Applied Statistics)} 69 (3):
547--64. \url{https://doi.org/10.1111/rssc.12401}.

\leavevmode\vadjust pre{\hypertarget{ref-heckman1976}{}}%
Heckman, James J. 1976. {``The {Common Structure} of {Statistical
Models} of {Truncation}, {Sample Selection} and {Limited Dependent
Variables} and a {Simple Estimator} for {Such Models}.''} In
\emph{Annals of {Economic} and {Social Measurement}, {Volume} 5, Number
4}, 475--92. {NBER}.

\leavevmode\vadjust pre{\hypertarget{ref-herrett2015}{}}%
Herrett, Emily, Arlene M Gallagher, Krishnan Bhaskaran, Harriet Forbes,
Rohini Mathur, Tjeerd van Staa, and Liam Smeeth. 2015. {``Data {Resource
Profile}: {Clinical Practice Research Datalink} ({CPRD}).''}
\emph{International Journal of Epidemiology} 44 (3): 827--36.
\url{https://doi.org/10.1093/ije/dyv098}.

\leavevmode\vadjust pre{\hypertarget{ref-higgins2009}{}}%
Higgins, Julian P. T., Simon G. Thompson, and David J. Spiegelhalter.
2009. {``A Re-Evaluation of Random-Effects Meta-Analysis.''}
\emph{Journal of the Royal Statistical Society: Series A (Statistics in
Society)} 172 (1): 137--59.
\url{https://doi.org/10.1111/j.1467-985X.2008.00552.x}.

\leavevmode\vadjust pre{\hypertarget{ref-little1996}{}}%
Little, Roderick J. A., and Yongxiao Wang. 1996. {``Pattern-{Mixture
Models} for {Multivariate Incomplete Data} with {Covariates}.''}
\emph{Biometrics} 52 (1): 98. \url{https://doi.org/10.2307/2533148}.

\leavevmode\vadjust pre{\hypertarget{ref-liu2021}{}}%
Liu, Dawei, Hanne I. Oberman, Johanna Muñoz, Jeroen Hoogland, and Thomas
P. A. Debray. 2021. {``Quality Control, Data Cleaning, Imputation.''}
In. {arXiv}. \url{https://arxiv.org/abs/2110.15877}.

\leavevmode\vadjust pre{\hypertarget{ref-morris2019}{}}%
Morris, Tim P., Ian R. White, and Michael J. Crowther. 2019. {``Using
Simulation Studies to Evaluate Statistical Methods.''} \emph{Statistics
in Medicine} 38 (11): 2074--2102.
\url{https://doi.org/10.1002/sim.8086}.

\leavevmode\vadjust pre{\hypertarget{ref-ogundimu_collins19}{}}%
Ogundimu, Emmanuel O., and Gary S. Collins. 2019. {``A Robust Imputation
Method for Missing Responses and Covariates in Sample Selection
Models.''} \emph{Statistical Methods in Medical Research} 28 (1):
102--16. \url{https://doi.org/10.1177/0962280217715663}.

\leavevmode\vadjust pre{\hypertarget{ref-thedhsprogram2020}{}}%
Program, The DHS. 2020. {``{DHS Survey Design}: {Malaria
Parasitemia}.''} {USA}: {U.S. Agency for International Development
(USAID)}.

\leavevmode\vadjust pre{\hypertarget{ref-puhani1997}{}}%
Puhani, Patrick A. 1997. {``Foul or {Fair}? {The Heckman Correction} for
{Sample Selection} and {Its Critique}. {A Short Survey}.''} 97-07.
{Leibniz, Germany}: {ZEW - Leibniz Centre for European Economic
Research}.

\leavevmode\vadjust pre{\hypertarget{ref-Rpackage}{}}%
R Core Team. 2021. \emph{R: A Language and Environment for Statistical
Computing}. Vienna, Austria: R Foundation for Statistical Computing.
\url{https://www.R-project.org/}.

\leavevmode\vadjust pre{\hypertarget{ref-radiceGJRMGeneralisedJoint2021}{}}%
Radice, Giampiero Marra and Rosalba. 2021. {``{GJRM}: {Generalised Joint
Regression Modelling}.''} https://CRAN.R-project.org/package=GJRM.

\leavevmode\vadjust pre{\hypertarget{ref-resche-rigon2018}{}}%
Resche-Rigon, Matthieu, and Ian R White. 2018. {``Multiple Imputation by
Chained Equations for Systematically and Sporadically Missing Multilevel
Data.''} \emph{Statistical Methods in Medical Research} 27 (6):
1634--49. \url{https://doi.org/10.1177/0962280216666564}.

\leavevmode\vadjust pre{\hypertarget{ref-resche-rigon2013}{}}%
Resche-Rigon, Matthieu, Ian R. White, Jonathan W. Bartlett, Sanne A. E.
Peters, Simon G. Thompson, and on behalf of the PROG-IMT Study Group.
2013. {``Multiple Imputation for Handling Systematically Missing
Confounders in Meta-Analysis of Individual Participant Data.''}
\emph{Statistics in Medicine} 32 (28): 4890--4905.
\url{https://doi.org/10.1002/sim.5894}.

\leavevmode\vadjust pre{\hypertarget{ref-rubin76}{}}%
Rubin, Donald B. 1976. {``Inference and Missing Data.''}
\emph{Biometrika} 63 (3): 581--92.
\url{https://doi.org/10.1093/biomet/63.3.581}.

\leavevmode\vadjust pre{\hypertarget{ref-rubin1987}{}}%
---------. 1987. \emph{Multiple {Imputation} for {Nonresponse} in
{Surveys}}. Wiley {Series} in {Probability} and {Statistics}. {Hoboken,
NJ, USA}: {John Wiley \& Sons, Inc.}
\url{https://doi.org/10.1002/9780470316696}.

\leavevmode\vadjust pre{\hypertarget{ref-rugnao2019}{}}%
Rugnao, Sheila, Samuel Gonahasa, Catherine Maiteki-Sebuguzi, Jimmy
Opigo, Adoke Yeka, Agaba Katureebe, Mary Kyohere, et al. 2019. {``{LLIN
Evaluation} in {Uganda Project} ({LLINEUP}): Factors Associated with
Childhood Parasitaemia and Anaemia 3 Years After a National Long-Lasting
Insecticidal Net Distribution Campaign: A Cross-Sectional Survey.''}
\emph{Malaria Journal} 18 (1): 207.
\url{https://doi.org/10.1186/s12936-019-2838-3}.

\leavevmode\vadjust pre{\hypertarget{ref-simmonds2005}{}}%
Simmonds, Mark C, Julian P T Higginsa, Lesley A Stewartb, Jayne F
Tierneyb, Mike J Clarke, and Simon G Thompson. 2005. {``Meta-Analysis of
Individual Patient Data from Randomized Trials: A Review of Methods Used
in Practice.''} \emph{Clinical Trials} 2 (3): 209--17.
\url{https://doi.org/10.1191/1740774505cn087oa}.

\leavevmode\vadjust pre{\hypertarget{ref-smith03}{}}%
Smith, Murray D. 2003. {``Modelling Sample Selection Using {Archimedean}
Copulas.''} \emph{The Econometrics Journal} 6 (1): 99--123.
\url{https://doi.org/10.1111/1368-423X.00101}.

\leavevmode\vadjust pre{\hypertarget{ref-staedke2019}{}}%
Staedke, Sarah G., Moses R. Kamya, Grant Dorsey, Catherine
Maiteki-Sebuguzi, Samuel Gonahasa, Adoke Yeka, Amy Lynd, Jimmy Opigo,
Janet Hemingway, and Martin J. Donnelly. 2019. {``{LLIN Evaluation} in
{Uganda Project} ({LLINEUP}) \textendash{} {Impact} of Long-Lasting
Insecticidal Nets with, and Without, Piperonyl Butoxide on Malaria
Indicators in {Uganda}: Study Protocol for a Cluster-Randomised
Trial.''} \emph{Trials} 20 (1): 321.
\url{https://doi.org/10.1186/s13063-019-3382-8}.

\leavevmode\vadjust pre{\hypertarget{ref-staedke2021}{}}%
Staedke, Sarah, Martin Donnelly, Janet Hemingway, Moses Kamya, and Grant
Dorsey. 2021. {``{ClinEpiDB. Study: {LLINEUP Cluster Randomized
Trial}}.''}
(https://clinepidb.org/ce/app/workspace/analyses/DS\_7c4cd6bba9/new/details).

\leavevmode\vadjust pre{\hypertarget{ref-theemergingriskfactorscollaboration2007}{}}%
The Emerging Risk Factors Collaboration. 2007. {``The {Emerging Risk
Factors Collaboration}: Analysis of Individual Data on Lipid,
Inflammatory and Other Markers in over 1.1 Million Participants in 104
Prospective Studies of Cardiovascular Diseases.''} \emph{European
Journal of Epidemiology} 22 (12): 839--69.
\url{https://doi.org/10.1007/s10654-007-9165-7}.

\leavevmode\vadjust pre{\hypertarget{ref-vella1998}{}}%
Vella, Francis. 1998. {``Estimating {Models} with {Sample Selection
Bias}: {A Survey}.''} \emph{The Journal of Human Resources} 33 (1): 127.
\url{https://doi.org/10.2307/146317}.

\leavevmode\vadjust pre{\hypertarget{ref-viechtbauer2005}{}}%
Viechtbauer, Wolfgang. 2005. {``Bias and {Efficiency} of {Meta-Analytic
Variance Estimators} in the {Random-Effects Model}.''} \emph{Journal of
Educational and Behavioral Statistics} 30 (3): 261--93.
\url{https://doi.org/10.3102/10769986030003261}.

\leavevmode\vadjust pre{\hypertarget{ref-wojtys2018}{}}%
Wojtyś, Małgorzata, Giampiero Marra, and Rosalba Radice. 2018. {``Copula
Based Generalized Additive Models for Location, Scale and Shape with
Non-Random Sample Selection.''} \emph{Computational Statistics \& Data
Analysis} 127 (November): 1--14.
\url{https://doi.org/10.1016/j.csda.2018.05.001}.

\end{CSLReferences}

\bibliographystyle{unsrt}
\bibliography{bibfile.bib}

\end{document}